\documentclass[twocolumn,prx,reprint,notitlepage,superscriptaddress,floatfix]{revtex4}
\usepackage{lmodern}
\usepackage[latin9]{inputenc}
\usepackage{color}
\usepackage{bm}
\usepackage{amsmath}
\usepackage{amssymb}
\usepackage{graphicx}
\usepackage{hyperref}

\makeatletter

\providecommand{\tabularnewline}{\\}

\@ifundefined{textcolor}{}
{%
 \definecolor{BLACK}{gray}{0}
 \definecolor{WHITE}{gray}{1}
 \definecolor{RED}{rgb}{1,0,0}
 \definecolor{GREEN}{rgb}{0,1,0}
 \definecolor{BLUE}{rgb}{0,0,1}
 \definecolor{CYAN}{cmyk}{1,0,0,0}
 \definecolor{MAGENTA}{cmyk}{0,1,0,0}
 \definecolor{YELLOW}{cmyk}{0,0,1,0}
}

\usepackage{bm}
\usepackage{subfigure}

\newcommand{\beq}{\begin{equation}}
\newcommand{\eeq}{\end{equation}}
\newcommand{\bea}{\begin{eqnarray}}
\newcommand{\eea}{\end{eqnarray}}
\newcommand{\be}{\begin{equation}}
\newcommand{\ee}{\end{equation}}

\DeclareMathOperator{\Tr}{Tr}

\makeatother

\begin{document}
\title{Topological and nematic superconductivity mediated by ferro-SU(4)
fluctuations in twisted bilayer graphene}
\author{Yuxuan Wang}
\affiliation{Department of Physics, University of Florida, Gainesville, FL 32601}
\author{Jian Kang}
\affiliation{School of Physical Science and Technology \& Institute for Advanced
Study, Soochow University, Suzhou, 215006, China}
\author{Rafael M. Fernandes}
\affiliation{School of Physics and Astronomy, University of Minnesota, Minneapolis,
55455 MN}
\begin{abstract}
We propose an SU(4) spin-valley-fermion model to investigate the superconducting
instabilities of twisted bilayer graphene (TBG). In this approach,
bosonic fluctuations associated with an emergent SU(4) symmetry, corresponding
to combined rotations in valley and spin spaces, couple to the low-energy
fermions that comprise the flat bands. These fluctuations are peaked
at zero wave-vector, reflecting the ``ferromagnetic-like'' SU(4)
ground state recently found in strong-coupling solutions of microscopic
models for TBG. Focusing on electronic states related to symmetry-imposed
points of the Fermi surface, dubbed here ``valley hot-spots" and
``van Hove hot-spots", we find that the coupling to the itinerant electrons
partially lifts the huge degeneracy of the ferro-SU(4) ground state
manifold, favoring inter-valley order, spin-valley coupled order,
ferromagnetic order, spin-current order, and valley-polarized order, depending on  details of the band structure. These fluctuations, in
turn, promote attractive pairing interactions in a variety of closely
competing channels, including a nodeless $f$-wave state, a nodal
$i$-wave state, and topological $d+id$ and $p+ip$ states with unusual Chern
numbers $2$ and $4$, respectively. Nematic superconductivity, although not realized
as a primary instability of the system, %can still 
appears as a consequence
of the near-degeneracy of superconducting order parameters that transform
as one-dimensional and two-dimensional irreducible representations
of the point group $D_{6}$.
\end{abstract}
\date{\today}
\maketitle

\section{Introduction}

The recent discovery of correlated insulating and superconducting
phases in twisted bilayer graphene (TBG) near the magic angle has
brought this system in the limelight of condensed matter physics~\cite{Pablo1,Pablo2,David,Young,Cory1,Cory2,Dmitry1,Ashoori,Dmitry2,Yazdani,Eva,Yazdani2,Shahal,Young2}.
Similar phenomena have also been identified in other moir\'e systems,
such as twisted double bilayer graphene~\cite{Pablo3,Guanyu,Kim},
twisted trilayer systems~\cite{Feng,Feng2,Feng3,KeWang19}, and transition metal dichalcogenide
moir\'e heterostructures \cite{Shan20}, illustrating that the observed
correlated electronic phases are rather universal in moir\'e systems.
In addition to the insulating and superconducting phases, electronic
nematic order \cite{Cory2,Perge,Eva,Cao20} and ferromagnetism \cite{David,Young,Dmitry1,YoungFM20}
have also been observed in TBG. Since many of these correlated phases
are also observed in unconventional superconductors, it is interesting
to compare the role of electronic correlations in these different
systems~\cite{Leon1,LiangPRX1,KangVafekPRX,Senthil1,FanYang,Louk,LiangPRX2,GuineaPNAS,Kivelson,Fernandes1,Guo,Kuroki,Qianghua,Stauber,KangVafekPRL,Bruno,Senthil2,Ashvin1,Cenke,MacDonald,Sachdev18,Zalatel1,Zalatel2,Guinea20,Senthil3Ferro,Sau,Ashvin2,Zalatel3,Dai2,YiZhang,Kang3,Chichinadze20,ZYMeng20,You,Rossi,Bernevig20,Torma,Ashvin20,DasSarma,Zaletel2005,Debanjan20,Balents20,SachdevPNAS,MacDonald20,Nandkishore219,Adam}.
At first sight, this seems to be a formidable task, because the tiny twist angles of TBG lead to a huge unit cell containing more than
$10^{4}$ atoms. However, it is widely understood now that the correlated
phases in TBG arise from the narrow bands around the charge neutrality
point (CNP), which are separated from the remote bands by a band gap
of tens of meV \cite{Bistritzer11,Mele11}. Such an observation suggests
that low-energy models focusing on the four narrow bands can shed
important light on the properties of TBG.

Due to the small bandwidth of these narrow bands, the kinetic energy
is comparable to the Coulomb interaction, pointing to the crucial
role played by electronic correlations in shaping the phase diagram
of TBG. Because the 
the Coulomb repulsion, estimated as $e^2/(\epsilon_{hBN} L_m) \approx 24$meV, is comparable with the calculated bandwidth \cite{Bistritzer11}, there have been parallel efforts on analyzing the system theoretically from both a strong-coupling
\cite{Leon1,LiangPRX1,Senthil1,Louk,KangVafekPRL,Kivelson,Fernandes1,Zalatel3,Kang3,Ashvin20}
and a weak-coupling perspective \cite{MacDonald,LiangPRX2,Nandkishore19,Bitan2019,Dai2,Guinea20,YiZhang,DasSarma20,Chichinadze20,LinNandkishore20}.
On the one hand, correlated insulating phases are experimentally observed
at commensurate fillings of the moir\'e superlattice \cite{Pablo1,Cory1,Dmitry1,Shahal,Yazdani2},
highlighting the importance of strong correlations. On the other hand,
van Hove singularities that occur at specific concentrations have
been proposed to host a variety of weak-coupling instabilities \cite{LiangPRX2,DasSarma20,Chichinadze20},
some of which may have been observed experimentally \cite{Cao20}.
In either approach, the correlated insulating state often breaks a
symmetry of the system. While it remains unsettled which -- if any
-- symmetries are broken in the correlated states of TBG, the fact
that superconductivity appears once the correlated state is suppressed
suggests %the possibility 
that fluctuations associated with the broken
symmetry of the correlated state may be responsible for the formation
of the Cooper pairs. Such a scenario parallels others widely employed
to model unconventional superconductors, such as pairing in the vicinity
of an antiferromagnetic state \cite{Scalapino12}. Of course, it is
possible that the superconductivity in TBG is a standard electron-phonon
pairing state, as proposed elsewhere \cite{Wu18,Bernevig19,Fabrizio19}.
Recent experiments also observed superconductivity in devices where
the strength of the correlations is suppressed, e.g. by decreasing
the distance between the metallic gates, by moving the system away
from the magic twist angle \cite{Dmitry2,Young2,Harpreet20}. Whether this is
an indication that pairing does not require correlations or that fluctuations
associated with the correlated state persist even after the latter
is suppressed remains under debate. 

\begin{figure}%[htp]
\includegraphics[width=0.8\columnwidth]{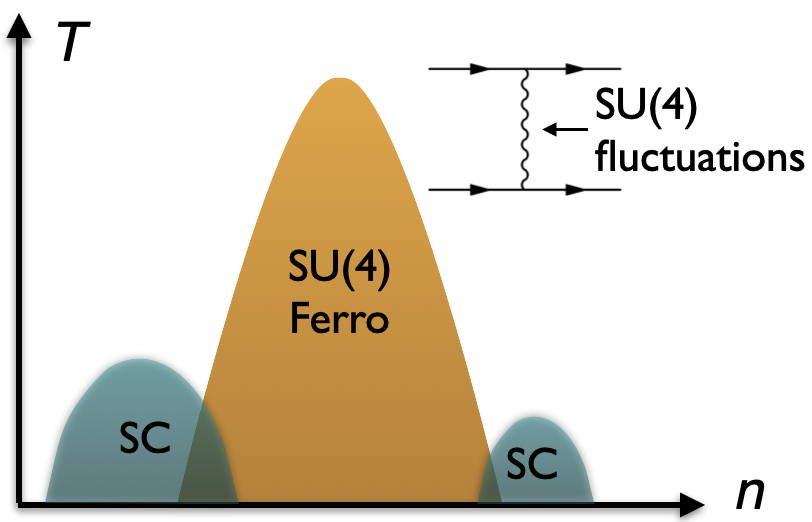}
\caption{Schematic phase diagram of superconductivity mediated by fluctuations associated with a correlated state that displays ``ferromagnetic-like" SU(4) order. As we show in this paper, the types of ferro-SU(4) order favored by the coupling to itinerant fermions are inter-valley order, spin-valley coupled order, ferromagnetic order, spin-current order, and valley-polarized order. As shown in Table \ref{tab:summary}, these fluctuations promote a rich landscape of pairing states,  such as a nodeless $f$-wave state, a nodal
$i$-wave state, and topological $d+id$ and $p+ip$ states with unusual Chern
numbers $2$ and $4$, respectively. }%\textcolor{magenta}{\textbf{RMF: I suggest we make this figure more TBG-like. Instead of x, the horizontal axis should be n. And the ferro-SU(4) order could be a full dome, with two different smaller SC domes (of different sizes) on the left and on the right of the ferro-SU(4) dome.}}}
\label{fig:phase}
\end{figure}

In this paper, we investigate the scenario in which superconductivity
in TBG arises from the fluctuations associated with the suppressed
correlated ordered state. We illustrate this scenario for superconductivity in the schematic phase diagram shown in Fig.~\ref{fig:phase}.
Instead of choosing between
a strong-coupling or a weak-coupling approach, we attempt to bridge
them by adopting a more phenomenological approach, similar in spirit
to the spin-fermion models widely employed to study magnetically-mediated
superconductivity in cuprates and heavy fermions \cite{Abanov03,Roussev01,Metlitski10,Wang13,Wang17}.
The traditional spin-fermion model is rooted in a separation of energy
scales: high-energy states give rise to an antiferromagnetic (or ferromagnetic)
SU(2) order parameter, while low-energy states interact with each
other via the exchange of fluctuations associated with this order
parameter. In this type of model, antiferromagnetic (ferromagnetic)
fluctuations are known to promote unconventional $d$-wave ($p$-wave)
pairing. It has been recently generalized to consider more generic
bosonic excitations, such as nematic \cite{Lederer15,Kang16,Klein19}
and ferroelectric fluctuations \cite{KoziiFu,WangCho,Gastiasoro20}, which
favor multiple pairing states simultaneously.

In TBG, the existence of well-separated energy scales is not as obvious
as in other quantum materials. Yet, given the difficulties in building
microscopic Hamiltonians for TBG and other moir\'e systems, and the powerful
insights that boson-fermion types of models have provided to other
quantum materials, it is interesting to build such a model for TBG
and investigate its predictions for the possible superconducting states.
In this regard, one of the most striking differences between TBG and
other unconventional superconductors is that, in the former, the fermions
are labeled by both spin and valley degrees of freedom. Because of
the negligibly small spin-orbit coupling of graphene, SU(2) spin-rotational
invariance is preserved. Moreover, TBG has an approximate U(1) valley
symmetry, related to the suppressed coupling between the two opposite
valleys of the graphene sheets for small twist angles \cite{Senthil1}.
Consequently, a sensible starting point is a low-energy model with
(approximate) SU(2)$\otimes$SU(2)$\otimes$U(1) symmetry, where the
two SU(2) groups correspond to independent spin rotations on the two
valleys. Because this is a subgroup of SU(4), it is often convenient
to label the possible ordered states in terms of a maximally symmetric
bosonic SU(4) order parameter $\hat{\Phi}$, related to rotations
in the combined spin and valley spaces, and thus characterized by
$15$ independent components \cite{Leon1,Fernandes1,Classen19,Natori19,KangVafekPRL}.
Interestingly, several low-energy microscopic Hamiltonians have been
proposed for TBG whose interacting parts display an emergent SU(4)
symmetry at the energy scale of the flat bands, which is generally
broken by the kinetic part. Strong coupling analyses have shown that
the ground state in this SU(4) manifold is ``ferromagnetic-like,''
i.e. it is described by an order parameter $\hat{\Phi}$ that condenses
at zero-momentum, without breaking translational symmetry \cite{KangVafekPRL,Bruno}.
Recent Quantum Monte Carlo simulations also found ferro-SU(4) ground
states \cite{ZYMeng20}. Experimentally, this is consistent with the
observation of ferromagnetism in TBG at certain fillings \cite{David,Young,YoungFM20}.
Importantly, the emergence at strong coupling of a ferromagnetic-like state, as opposed
to the more usual antiferromagnetic-like state, is closely
tied to the fragile topology of the flat bands \cite{KangVafekPRL,Zalatel1,Zalatel2,Senthil3Ferro,Kang3}.
Whether such a ferro-SU(4) state can be obtained directly within a
weak-coupling low-energy theory remains to be seen. Interestingly, the mean-field analysis of a related model that has SU(2) spin-rotational symmetry and U(1) orbital symmetry found a ``ferromagnetic-like" ground state, which can be either orbital nematic or orbital ferromagnetic \cite{Kivelson}. This orbitally ordered state forms a dome in the temperature-doping phase diagram, giving rise to two smaller spin-singlet orbital-triplet superconducting domes at its edges, similarly to our sketch in Fig. \ref{fig:phase}. As we will show below, our analysis, which includes fluctuations of the ordered state, also find regimes of spin-singlet valley-triplet superconductivity. 

The goal of our phenomenological model is to provide a useful platform to analyze the effects of strong coupling physics on the itinerant fermions. Denoting the low-energy fermionic creation operators by $\psi_{\alpha a}^{\dagger}$,
with $\alpha=\pm1\equiv\pm K$ referring to the valley degrees of
freedom and $a=\uparrow,\downarrow$, to the spin degrees of freedom,
the coupling between the boson $\hat{\Phi}$ and the low-energy fermions
is given by: 
\begin{equation}
H_{\mathrm{int}}=g\phi_{\mu\nu}\,\psi_{\alpha a}^{\dagger}\tau_{\alpha\beta}^{\mu}\sigma_{ab}^{\nu}\psi_{\beta b},\label{H_int}
\end{equation}
where $g$ is a coupling constant and the Pauli matrices $\boldsymbol{\tau}$
and $\boldsymbol{\sigma}$ act on valley space and on spin space,
respectively (summation over repeated indices is left implicit). Here,
$\phi_{\mu\nu}$ are the 15 components of $\hat{\Phi}$ with $\mu,\,\nu=0,...,\,3$
subject to the constraint that $\mu=\nu=0$ is excluded; this last
condition arises because $\phi_{00}$ is just a constant that can
be absorbed in the definition of the chemical potential. We dub this
phenomenological low-energy model the \emph{SU(4) spin-valley-fermion model}.

\begin{table*}
\centering \caption{A summary of the superconducting instabilities mediated by the exchange
of ferro-SU(4) fluctuations in different channels (first and second
columns). The pairing nomenclature in the third column ($s,\,p,\,d$,
etc) is explained in the main text in terms of the irreducible representations
of the $D_{6}$ group. The Chern number of each pairing state in the
third column is respectively shown in the fourth column. The last
column shows what combination of pairing states generate a subsidiary
nematic order.}
\begin{tabular}{|c|c|c|c|c|}
\hline 
SU(4) channel & SU(4) ground state & Pairing & Chern number & Nematicity\tabularnewline
\hline 
$\phi_{v}$ & inter-valley & valley-triplet: $A_{1}$ ($s$-wave), $E_{2}$ ($d\pm id$), $E_{1}$
($p\pm ip$) & 0, 4, 2 & $s\pm d$\tabularnewline
$\phi_{sv}$ & spin-valley & spin-singlet \& valley-singlet: $E_{2}$ ($d$$\pm id$), $A_{2}$ ($i$-wave) & 2, N/A & $i\pm d$\tabularnewline
$\phi_{s}$ & ferromagnetic & spin-triplet: $B_{2}$ ($f$-wave), $E_{1}$ ($p\pm ip$) & 0, 4 & $p\pm f$\tabularnewline
$\phi_{sc}$ & spin-current & spin-singlet: $A_{1}$ ($s$-wave), $E_{2}$ ($d\pm id$) & 0, 2 & $s\pm d$
\tabularnewline
$\phi_{vp}$ & valley-polarized &{no Cooper instability} & N/A & N/A 
\tabularnewline
\hline 
\end{tabular}\label{tab:summary}
\end{table*}

Here, we investigate the superconducting (SC) states --- and their
accompanying topological and lattice symmetry-breaking properties
--- that are mediated by the exchange of ferro-SU(4) fluctuations
between low-energy fermions. In a single-band metal, it is well known
that ferromagnetic SU(2) fluctuations cause repulsion in the conventional
$s$-wave spin-singlet pairing channel and attraction in the $p$-wave
spin-triplet channel \cite{Roussev01}. In our case, because of the
multi-flavor nature of the fermions and of the much larger SU(4) manifold
of the bosonic field, the model has a richer structure in terms of
bosonic fluctuations and pairing channels \cite{Fernandes1,Scheurer20,Sachdev20}.
In particular, because the dispersion of the low-energy fermions explicitly
breaks the emergent SU(4) symmetry, we find that the boson-fermion
coupling of Eq.~(\ref{H_int}) lifts the SU(4) degeneracy and restricts
the soft fluctuations to a few distinct channels, namely: the
ferromagnetic channel, characterized by $\phi_{s}\equiv\phi_{0i}$;
the spin-current channel, given by $\phi_{sc}\equiv\phi_{3i}$; the
inter-valley channel, described by $\phi_{v}\equiv\phi_{(1,2)0}$; the valley-polarization channel, described by $\phi_{vp}\equiv\phi_{30}$;
and the spin-valley coupled channel, associated with $\phi_{sv}\equiv\phi_{(1,2)i}$.
Despite the fact that the detailed band structure and the corresponding
Fermi surface (FS) of TBG remain under debate, we argue that the selection
of the strongest SU(4)-fluctuations channels, as well as of the corresponding
leading pairing instabilities, depend only on robust features of the
FS. In particular, when the FS of a given valley is close to a van
Hove singularity~\cite{LiangPRX2,FuNC,Chichinadze19}, the coupling
to the so-called van Hove hot-spots (see Fig. \ref{fig:1}(b)) enhances
the bosonic fluctuations in the intra-valley spin ferromagnetic channel
($\phi_{s}$), the spin current channel ($\phi_{sc}$), and the valley-polarized channel ($\phi_{vp}$). On the other
hand, the coupling to states close to the intersections of the FSs
formed by the two valleys (called valley hot-spots, see Fig. \ref{fig:1}(a))
strongly enhances fluctuations in the inter-valley channel, which
can be either in the density sector ($\phi_{v}$) or in the the spin
sector ($\phi_{sv}$).

A comparative analysis of the strength
of these different SU(4)-fluctuations channels requires microscopic details of the band structure beyond the scope of this work. Instead, we analyze the pairing
states induced by each channel separately, discussing their implications
to the phase diagram of TBG. A crucial point is that the pairing symmetry,
as well as the topology of the induced SC order by each $\phi_{\mu\nu}$,
can be obtained without detailed knowledge of the FS, by linearizing
the low-energy fermionic dispersions around the van Hove and valley
hot-spots. Due to the multi-flavor nature of the fermions and the
ferro-like nature of the SU(4) fluctuations, which are peaked at zero
momentum, we find that for all SU(4) channels considered, multiple
pairing channels are simultaneously enhanced. Qualitative energetic
arguments are then made to determine the leading pairing channels.
Our main results are summarized in Table \ref{tab:summary}, which
displays the leading pairing states favored by the different SU(4)-fluctuations
channels discussed above. As indicated by their Chern numbers, several
of these states are topologically non-trivial superconductors. Moreover,
because multiple orthogonal pairing channels are simultaneously favored
by a given sector of SU(4) fluctuations, this opens up the possibility
of accidental degeneracy between different superconducting states.
In the vicinity of such an accidental degeneracy, the sixfold rotational
symmetry of the moir\'e lattice can be broken. The latter case gives rise
to an accompanying electronic nematic order inside the superconducting
state, as recently observed experimentally \cite{Cao20}.

The remainder of the paper is organized as follows. In Sec.~\ref{sec:model}
we present the SU(4) spin-valley-fermion model, and in Sec.~\ref{sec:fluctuations}
we discuss the lifting of the maximal SU(4) symmetry in the normal
state. In Sec.~\ref{sec:pairing} we analyze the leading pairing
channels induced by various SU(4) fluctuation channels and their symmetry-breaking
and topological properties. In Sec.~\ref{sec:nem} we show that in
all channels considered, the near-degenerate pairing channels can
further induce an additional nematic order parameter. Conclusions
are presented in Sec. \ref{sec:Summary}.

\section{The SU(4) spin-valley-fermion model}

\label{sec:model}

We consider itinerant fermions coupled to soft SU(4) fluctuations
peaked at zero momentum, i.e. ``ferro-SU(4)'' fluctuations. In each
moir\'e unit cell, there are three relevant fermionic degree of freedom
-- moir\'e sublattice, spin, and valley --- giving rise to four spin-degenerate
nearly-flat bands. Here, ``valley'' refers to the valleys of the
underlying graphene atomic lattice and not to the valleys of the moir\'e
Brillouin zone. If the twist angle is small, electrons
with different valley indices do not mix, and one can approximately
treat valley as a good quantum number. Thus each low-energy fermion
can be labeled by its valley and spin quantum numbers.

The action of the SU(4) spin-valley-fermion model is written as
\begin{align}
 & \mathcal{S}=\int d\bm{k}d\omega\left[\Psi^{\dagger}(\omega,\bm{k})(-i\omega+\hat{\epsilon}(\bm{k}))\Psi(\omega,\bm{k})\right.\nonumber \\
 & +\left.g\Psi^{\dagger}(\omega,\bm{k})\hat{\Phi}(\omega-\omega',\bm{k}-\bm{k}')\Psi(\omega',\bm{k}')\right]\nonumber \\
 & +\frac{1}{2}\int d\bm{q}d\Omega\ (r+\alpha\Omega^{2}+\beta\bm{q}^{2})\Tr[\hat{\Phi}(\Omega,\bm{q})\hat{\Phi}(\Omega,\bm{q})],\label{eq:1}
\end{align}
where $g$ is a coupling constant, $\Psi$ is a four-component spinor
that combines spin and valley degrees of freedom, and the bosonic
order parameter $\hat{\Phi}$ is a four-by-four matrix that satisfies
$\Tr(\hat{\Phi})=0$. Notice that we work in the band basis and have
only kept the two bands (per spin) that give rise to Fermi surfaces
\cite{LiangPRX1,KangVafekPRX}. One can expand the field $\hat{\Phi}$
as a linear superposition of the 15 generators $\lambda_{i}$ of SU(4),
$\hat{\Phi}(\Omega,\bm{q})=\sum_{i=1}^{15}\phi_{i}(\Omega,\bm{q})\lambda_{i}$.
Without the kinetic energy term $\hat{\epsilon}(\bm{k})$, this model
has an SU(4) symmetry, under which the 15 bosonic fields $\phi_{i}$
transform in the adjoint representation. Here, the quantity $(r+\alpha\Omega^{2}+\beta\bm{q}^{2})$
in the last term is the (bare) inverse propagator of the bosonic fluctuations.
At the bare level, when $r$ becomes negative, the $\hat{\Phi}$ field
condenses at zero-momentum and the system enters a ferro-SU(4) ordered
state. Unlike the usual SU(2) spin-ferromagnetic states, the fermions
can order in the spin, valley, and spin-valley coupling channels,
and the ground state manifold has a much larger symmetry. 

For our
purposes, it is more convenient to reexpress the matrix field $\hat{\Phi}$
in terms of the tensor product of spin and valley Pauli matrices,
$\tau^{\mu}\otimes\sigma^{\nu}$:
\begin{equation}
\hat{\Phi}(\Omega,\bm{q})={\sum_{\mu,\nu=0}^{3}}{}'\,\phi_{\mu\nu}(\Omega,\bm{q})\tau^{\mu}\otimes\sigma^{\nu}
\end{equation}
where the summation $\sum'$ excludes the term $\mu=\nu=0$.

The SU(4) symmetry is broken explicitly by the free-fermion term of
the action (\ref{eq:1}). In particular, the fermionic dispersion
of the two valleys from the underlying atomic lattice (labeled by
$K$ and $K'=-K$) are distinct: 
\begin{equation}
\hat{\epsilon}(\bm{k})=\left[\frac{\epsilon_{K}(\bm{k})+\epsilon_{K'}(\bm{k})}{2}\tau^{0}+\frac{\epsilon_{K}(\bm{k})-\epsilon_{K'}(\bm{k})}{2}\tau^{3}\right]\otimes\sigma^{0}\label{eq:dispersion}
\end{equation}
where $\epsilon_{K}(\bm{k})\neq\epsilon_{K'}(\bm{k})$ and hereafter
we will omit the $\otimes$ sign for compactness. Due to the approximate
valley symmetry, the system transforms under the point group $D_{6}$.
The fermionic dispersions associated with the $K$ and $K'$ valleys
are invariant under the three-fold rotation around the $z$-axis,
$C_{3z}$, and the in-plane two-fold rotation $C_{2y}$. Additionally,
the two valleys swap under the other in-plane two-fold rotation $C_{2x}$
and the two-fold rotation around the $z$-axis $C_{2z}$, as well
as the $C_{6z}$ rotation. By these symmetry requirements, fermions
in each valley give rise to a Fermi surface (FS) that has the same
symmetry as a triangle, as illustrated in Fig.~\ref{fig:1}, and
the two FSs are related by $C_{2z}$ (or, alternatively, $C_{2x}$
and $C_{6z}$). The exact shape of the FSs will strongly depend on
the details of the microscopic hopping parameters, which remain unclear
at present. However, much of our analysis does not rely on such details.

It is instructive to analyze the transformation properties under $D_{6}$
and time-reversal symmetries in the valley subspace. The elements
in $D_{6}$ that act non-trivially in valley subspace are $C_{6}$,
$C_{2z}$, and $C_{2x}$, all of which exchange the two valleys and are represented by $\tau^1$. As a result, 
the valley matrix
$\tau^{1}$ transforms as $\tau^{1}\to C_{6}\tau^{1}C_6^{-1} \equiv \tau^1$, and
 thus belongs to the irreducible representation
(irrep) $A_{1}$ of $D_{6}$. By the same token, $\tau^{2}$
and $\tau^{3}$ belong to irrep $B_{2}$. Moreover,
time reversal $\mathcal{T}$ also swaps the two valleys, i.e., $\mathcal{T}=\tau^1K$, where $K$ is complex conjugation. Under time-reversal, $\tau^{1}$
and $\tau^{2}$ are even but $\tau^{3}$ is odd. We list the details
of these transformation properties in Table \ref{tab:irreps}. We
note that in the Wannier orbital basis in which the ferro-SU(4) state
is most intuitively derived, the point group symmetry is lowered to
$D_{3}$, unless one includes additional orbitals that form high-energy
bands. This is a result of the fragile topology of the flat bands
that dictates that the $C_{2z}$ symmetry cannot be implemented locally
in the Wannier basis \cite{Senthil1,KangVafekPRX}. Since we work
in the band basis, this issue is avoided. 
\begin{table}
\label{tab:irreps} \centering \caption{Transformation properties of the Pauli matrices in valley subspace
under the symmetry operations $C_{2z}$ and time reversal $\mathcal{T}$.}
\begin{tabular}{|c|c|c|c|}
\hline 
valley matrix  & irrep of $D_{6}$  & $C_{2z}=\tau^{1}$  & $\mathcal{T}=\tau^{1}K$ \tabularnewline
\hline 
$\tau^{1}$ & $A_{1}$  & 1  & 1 \tabularnewline
$\tau^{2}$  & $B_{2}$  & $-$1  & 1 \tabularnewline
$\tau^{3}$  & $B_{2}$  & $-$1  & $-$1 \tabularnewline
\hline 
\end{tabular}
\end{table}

As discussed in the Introduction, we take the existence of a ferro-SU(4)
ground state as an input of our model, motivated by strong-coupling
analyses, Quantum Monte Carlo simulations, and experimental results
showing yet no evidence of translational symmetry breaking inside
the insulating phases. In this sense, we assume a separation of energy
scales, such that strong-coupling effects give rise to a largely degenerate
ferro-SU(4) ground state. Upon suppression of the insulating ferro-SU(4)
ground state, which can be achieved by doping or by increasing the
separation between the metallic gates of TBG (which causes a suppression
of the Coulomb repulsion \cite{Dmitry2}), ferro-SU(4) fluctuations
persist and then couple to the low-energy fermions. Our goal is then
twofold: (i) to determine which SU(4) fluctuation channels, represented
by $\phi_{\mu \nu}$, are enhanced by the coupling to the fermions, as the
fermionic dispersion explicitly breaks the SU(4) symmetry; and (ii)
to determine which pairing states are favored by those enhanced SU(4)
fluctuations. These two problems are addressed in the upcoming Sections
\ref{sec:fluctuations} and \ref{sec:pairing}, respectively. The
system is strongly coupled when the SU(4) fluctuations are soft, i.e.,
when the mass term $r$ is renormalized to zero, potentially leading
to non-Fermi liquid and strange metal behaviors. However, in this
work we focus on the symmetries of the ferro-SU(4) fluctuations and
the types of pairing symmetries, leaving a detailed analysis of the
dynamics of the coupled boson-fermion system to future work.

\section{Lifting of SU(4) degeneracy}

\label{sec:fluctuations}

As discussed above, the kinetic part of the Hamiltonian, Eq.~(\ref{eq:dispersion}),
lifts the emergent SU(4) degeneracy of the interacting part. Despite
the lack of complete knowledge of $\epsilon_{K}(\bm{k})$ and $\epsilon_{K'}(\bm{k})$,
their effect on selecting particular SU(4)-fluctuations channels can
be deduced from the symmetry-imposed features of the Fermi surface
and from the coupling in Eq. (\ref{H_int}).

\begin{figure}
\subfigure[\label:]{\includegraphics[width=0.47\columnwidth]{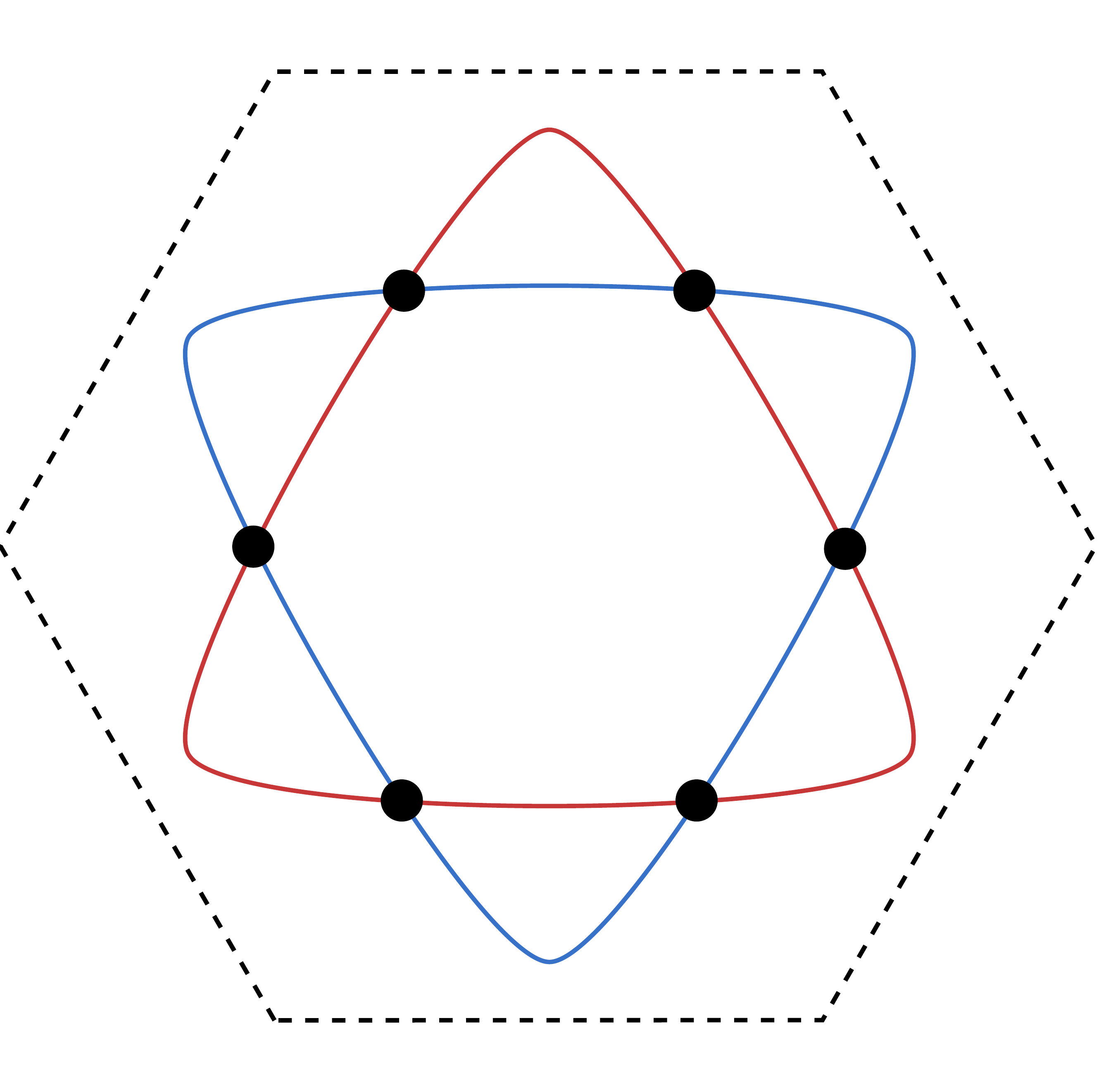}}~~~
\subfigure[\label:]{\includegraphics[width=0.47\columnwidth]{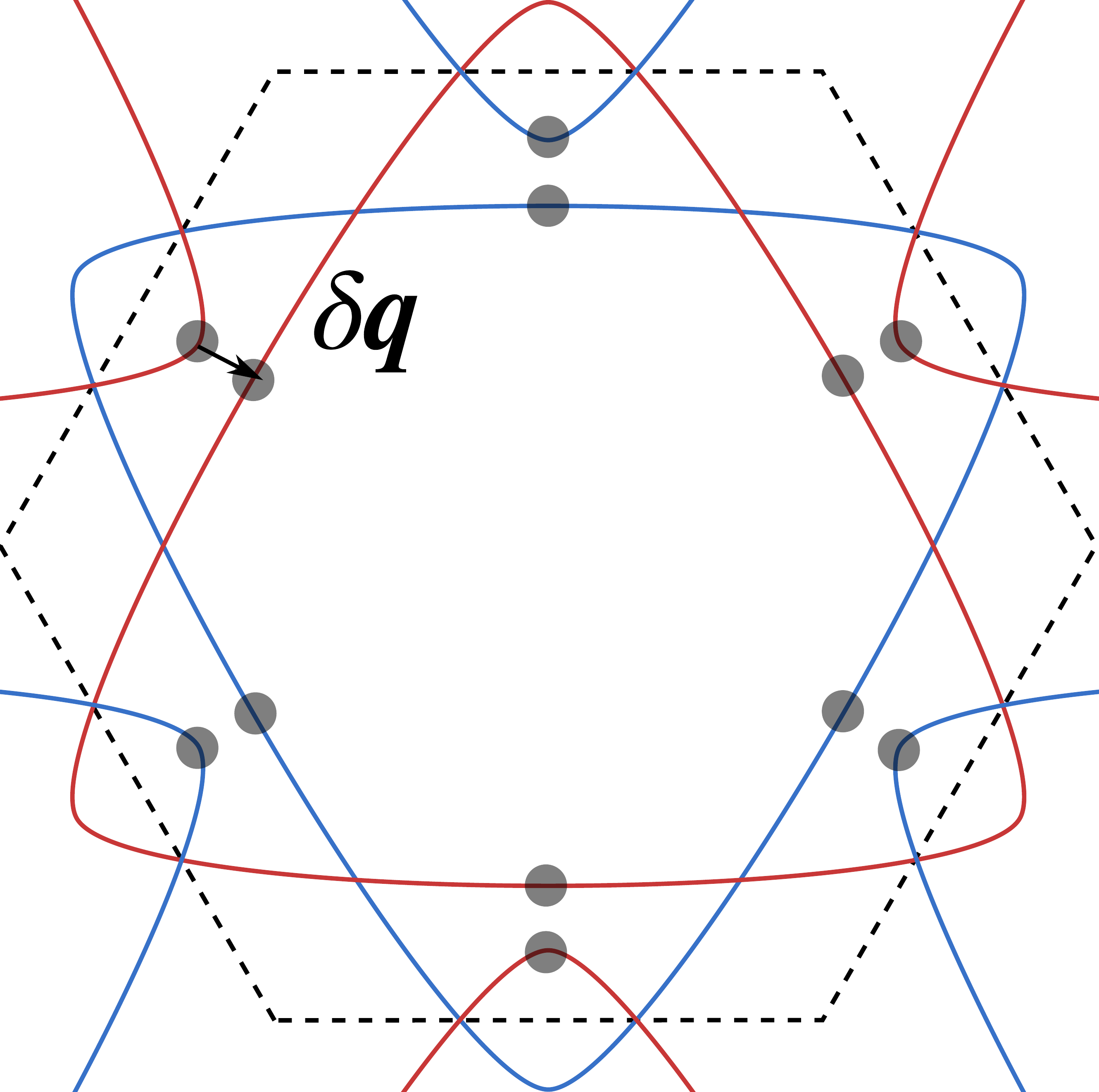}}
\caption{Schematics of the Fermi surface. Blue and red correspond to different
valleys. Regardless of band structure details, these Fermi surfaces
display two prominent features: the symmetry-imposed intersection
points between the Fermi surfaces from the two valleys (the valley
hot-spots, highlighted with black dots in panel (a)), and the proximity
to a van Hove singularity for appropriate doping levels (the van Hove
hot-spots, highlighted with grey dots in panel (b)). Valley hot-spots are located along the $\bar{\Gamma}-\bar{K}$ and $\bar{\Gamma}-\bar{K}'$ lines, whereas the van Hove hot-spots are located along the $\bar{\Gamma}-\bar{M}$ lines of the moir\'e Brillouin zone.}
\label{fig:1} 
\end{figure}

First, due to the approximate absence of inter-valley coupling, the
two sets of FSs formed by valleys $K$ and $K'$ intersect at points
that are invariant under the rotations $C_{2x}$ and $C_{2z}$. Following
the standard terminology, we refer to them as ``valley hot-spots''
of the Brillouin zone (BZ); as shown in Fig.~\ref{fig:1}(a), there
are six such valley hot-spots. In the vicinity of these points, the
fermions interact via exchanging low-energy ``inter-valley'' fluctuations,
which can be mediated by either of the two scalar fields $\phi_{v}\equiv\left(\phi_{10},\phi_{20}\right)$
or by either of the two vector fields $\phi_{sv}\equiv\left(\phi_{1i\neq0},\phi_{2i\neq0}\right)$.
Note that, from a point-group symmetry perspective (see Table \ref{tab:irreps}),
$\phi_{10}$ and $\phi_{20}$ transform as different irreps of $D_{6}$,
$A_{1}$ and $B_{2}$ respectively. Nevertheless, we keep them as
two components of the same field $\phi_{v}$, since they form a representation of the emergent valley U(1) symmetry, and because
they also promote the same pairing instabilities (see below).
Due to the valley U(1) symmetry, coupling
to the valley hot-spots does not lift their degeneracy. Moreover,
for commensurate twist angles, the underlying point group is explicitly
lowered to $D_{3}$, implying that both $\phi_{10}$ and $\phi_{20}$ transform
as the trivial irrep of $D_{3}$. %\textcolor{magenta}{\textbf{RMF Yuxuan, I think that incommensurate angles give D6, and commensurate angles give D3}} 
Similar
considerations apply to $\phi_{1i}$ and $\phi_{2i}$. We dub these
two sets of order parameters inter-valley order ($\phi_{v}$) and
spin-valley coupled order ($\phi_{sv}$), respectively. 

In the inter-valley channel, the $\phi_{v}$ boson couples to the
fermions via $g\phi_{v}^{1,2}\Psi^{\dagger}\tau_{1,2}\Psi$. This
coupling reduces the renormalized mass term $r$ of the SU(4) fluctuations
in Eq.~\eqref{eq:1}, but only in the $\phi_{v}$ channel. As a result,
this fermion-boson coupling term lifts the SU(4) degeneracy of the
fluctuations by enhancing the fluctuations in the inter-valley channel.
To see this, we can calculate the renormalized mass $r_{v}$ for the
inter-valley fluctuations $\phi_{v}$ within one-loop order: 
\begin{align}
r_{v} & =r+g^{2}\sum_{n,\omega_{m}}\int d\bm{k}\frac{1}{i\omega_{m}-{\epsilon}_{K}^{n}(\bm{k})}\frac{1}{i\omega_{m}-{\epsilon}_{K'}^{n}(\bm{k})}\nonumber \\
 & \approx r-g^{2}\sum_{n}\int\frac{d\epsilon_{1}d\epsilon_{2}[n_{F}(\epsilon_{2})-n_{F}(\epsilon_{1})]}{|\bm{v}_{K}^{n}\times\bm{v}_{K'}^{n}|(\epsilon_{2}-\epsilon_{1})}
\end{align}
where the summation $\sum_{n}$ is over all $n$ valley hot-spots,
and ${\epsilon}_{K'}^{n}(\bm{k})=\bm{v}_{K}^{n}\cdot\bm{k}$ is the
linear approximation of the dispersion around each hot spot. Here,
$n_{F}(\epsilon)$ is the Fermi-Dirac distribution function. The integral
requires an upper cutoff, but it is clear that the low-energy fermions
provide a downward renormalization to $r$ in the inter-valley channel.

At the same one-loop order, the mass of the spin-valley coupled fluctuations
also gets enhanced by the same mechanism. These fluctuations couple
to the same valley hot-spot via $\sum_{i}\ensuremath{\phi_{sv}^{i,1}\Psi^{\dagger}\sigma_{i}\tau_{1}\Psi+\phi_{sv}^{i,2}\Psi^{\dagger}\sigma_{i}\tau_{2}\Psi}$.
Since the dispersion is spin-degenerate, the renormalization of the
corresponding mass term $r_{sv}$ is the same as that for $r_{v}$.
Thus, the fluctuations corresponding to $\phi_{sv}$ also get enhanced.
At one-loop level, the inter-valley and spin-valley coupled fluctuations
are therefore degenerate. But such a degeneracy is not protected by
any symmetry, and can be lifted by other effects, such as small perturbations
at high energies. For this reason, we will consider $\phi_{v}$ and
$\phi_{sv}$ separately.

Aside from the generic considerations above, the proximity to van
Hove singularities (and higher-order van Hove singularities) of the
FS has been argued to play an important role, particularly for certain
electronic concentrations in the TBG phase diagram \cite{LiangPRX2,Nandkishore19,Chichinadze19,DasSarma20,Chichinadze20}.
The van Hove singularities considered in this work take place when
the FSs cross the Brillouin zone boundaries. Near the van Hove singularity,
thus, for a FS of a given valley, there are pairs of points separated
by a small momentum $\delta\bm{q}$ with parallel velocities and enhanced
density of states; right at the van Hove singularity, $\delta\bm{q}\to0$.
We denote these pairs of points van Hove hot-spots, as shown in Fig.~\ref{fig:1}(b).
The components of the SU(4) bosonic field $\hat{\Phi}$ that couple
to the van Hove hot-spots are the intra-valley ones, which form two
three-component fields $\phi_{s}\equiv\phi_{0,i\neq0}$ and $\phi_{sc}\equiv\phi_{3,i\neq0}$, as well as a single component field $\phi_{vf}\equiv\phi_{30}$.
Clearly, $\phi_{s}$ describes a ferromagnetic order parameter, and,
due to the additional valley component, $\phi_{sc}$ describes a spin-current
order parameter, breaking spin-rotation and $C_{2z}$ but not time-reversal
symmetry. On the other hand, $\phi_{vp}$ corresponds to a valley-polarization order, breaking $C_{2z}$ and time-reversal.

In the case of $\phi_{s}$, the van Hove hot-spots primarily exchange
(intra-valley) ferromagnetic spin fluctuations via the coupling term
$g\phi_{s}^{i}\Psi^{\dagger}\sigma_{i}\Psi$. Within one-loop, both
the mass term $r_{s}$ and the stiffness term $\beta_{s}\delta\bm{q}^{2}$
corresponding to $\phi_{s}$ are renormalized according to 
\begin{align}
r_{s}+\beta_{s}\delta\bm{q}^{2} & =r+\beta\delta\bm{q}^{2}\nonumber \\
 & +g^{2}\sum_{n,\omega_{m}}\int d\bm{k}\frac{1}{i\omega_{m}-{\epsilon}^{n}(\bm{k})}\frac{1}{i\omega_{m}-{\epsilon}^{n}(\bm{k}+\delta\bm{q})}\label{eq:phis}
\end{align}
where ${\epsilon}^{n}(\bm{k})$ corresponds to the dispersion expanded
around the $n$-th pair of van Hove hot-spots. Because the Fermi velocities
are opposite within a pair of van Hove hot-spots, the mass term $r_{s}$
and the stiffness $\beta_{s}$ are suppressed by the one-loop contribution.
At the van Hove singularity, $\delta\bm{q}\rightarrow0$, the density
of states diverges and the Fermi surface becomes non-analytic. The
precise form of the $\phi_{s}$ propagator depends on details of the
van Hove singularity, which we do not pursue here. In any case, due
to the enhanced density of states, near a van Hove singularity, one
expects that the enhancement of ferromagnetic fluctuations will be
larger than the enhancement in the inter-valley and spin-valley coupled channels
discussed above. We also note that the proximity to a van Hove singularity
on its own may favor other weak-coupling instabilities that can break
translational symmetries, as discussed e.g. in \cite{LiangPRX2,Chichinadze20};
here, our focus is instead on how the van Hove singularity lifts the
degeneracy of the SU(4) fluctuations that arise from strong-coupling
physics.

In the cases of $\phi_{sc}$ and $\phi_{vp}$, i.e., the spin-current and valley-polarized orders, fermions
interact with the bosons via the coupling $g\phi_{sc}^{i}\Psi^{\dagger}\tau_{3}\sigma_{i}\Psi$ and $g\phi_{vp}\Psi^{\dagger}\tau_{3}\Psi$.
Their corresponding mass terms get renormalized downwards in the same
way as in Eq.~\eqref{eq:phis}. As a result, $\phi_{sc}$ and $\phi_{vp}$ fluctuations
get enhanced in the same manner as those of $\phi_{s}$, at least at one-loop order. However,
because they are not related by any symmetry, effects
beyond one-loop can lift their degeneracy. For this reason we treat
them separately when analyzing the pairing instabilities.

\section{Pairing states mediated by SU(4) fluctuations \label{sec:pairing}}

Having established the dominant ferro-SU(4) fluctuation channels,
we now analyze the pairing states favored by the exchange of such
fluctuations between low-energy fermions. Fluctuations with small
momentum transfer can typically favor multiple superconducting orders
with different pairing symmetries; examples include the case of nematic
fluctuations~\cite{Lederer15,Kang16,Klein19}, ferroelectric fluctuations~\cite{KoziiFu,WangCho,Gastiasoro20},
and charge-order fluctuations with small momentum~\cite{WangChubukovCF}.
We will thus focus mainly on the internal structure of the pairing
state and determine its spatial dependence based solely on symmetry
and energetic argumens. Specifically, when multiple gap functions
solve the same gap equation, we will focus only on those gap structures
that are nodeless, since a fully gapped state is generally expected
to be energetically favored over a nodal state by maximizing the condensation
energy.

Generically, with spin and valley degrees of freedom, the SC order
parameter couples to fermions via $\hat{\Delta}\Psi^{\dagger}(\bm{k})\Psi^{\dagger T}(-\bm{k})$.
We can restrict the form of the possible SC orders by energetics and
symmetry arguments. The $D_{6}$ symmetry group acts on both momentum
space \emph{and} valley space; due to the absence of spin-orbit coupling,
it acts trivially on spin space. In particular, the two valleys $K$
and $K'$ are exchanged under $C_{2x}$ and $C_{2z}$ (see Table \ref{tab:irreps}).
Since low-energy fermions at $\bm{k}$ and $-\bm{k}$ necessarily
come from two different valleys, we consider Cooper pairing between
the two valleys only. As a result, the SC terms are inter-valley ones,
which can further be distinguished by ``parity'', i.e. the eigenvalues
of $C_{2z}$, which takes $\left(k_{x},k_{y}\right)$ to $\left(-k_{x},-k_{y}\right)$.
For an even-parity (spin-singlet) order parameter, the pairing vertex
is given by
\begin{align}
\hat{\Delta}=\left[\Delta_{o}(\bm{k})i\tau_{2}+\Delta_{e}(\bm{k})\tau_{1}\right](i\sigma_{2}),\label{eq:5}
\end{align}
and for an odd-parity (spin-triplet) order parameter 
\begin{align}
\hat{\Delta}'_{j}=\left[\Delta'_{e}(\bm{k})i\tau_{2}+\Delta'_{o}(\bm{k})\tau_{1}\right](\sigma_{j}\,i\sigma_{2}),\label{eq:5'}
\end{align}
where $\Delta_{e}$ and $\Delta'_{e}$ are even functions of momentum
while $\Delta_{o}$ and $\Delta'_{o}$ are odd. The $i\tau_{2}$ term,
which is odd under $C_{2z}$, corresponds to valley-singlet pairing,
whereas the $\tau_{1}$ term, which is even under $C_{2z}$, corresponds
to valley-triplet. Since the SU(2) valley symmetry is broken by the
FS, in general both terms appear.

Within the $D_{6}$ point group of TBG considered here, even-parity
superconducting order parameters must transform according to the irreducible
representations $A_{1},A_{2},E_{2}$, whereas odd-parity ones transform
according to the irreps $B_{1},B_{2},E_{1}$. The $A_{1,2}$ and $B_{1,2}$
irreps are one-dimensional while the $E_{1,2}$ irreps are two-dimensional.
It is common in the literature to associate a SC state transforming
as an irrep of a point group to an orbital angular momentum $s$,
$p$, $d$, etc. This is usually done by identifying the spherical
harmonic corresponding to the lowest-order basis function of that
irrep. Then, $A_{1}$ corresponds to $s$-wave, $A_{2}$ to $i$-wave,
$B_{1}$ to $f_{x\left(x^{2}-3y^{2}\right)}$-wave (hereafter denoted
$f'$-wave for simplicity), $B_{2}$ to $f_{y\left(3x^{2}-y^{2}\right)}$-wave
(hereafter denoted $f$-wave for simplicity), $E_{1}$ to $(p_{x},\,p_{y})$-wave,
and $E_{2}$ to $(d_{x^{2}-y^{2}},\,d_{xy})$-wave.

The corresponding irrep of an order parameter of the form of Eqs.~(\ref{eq:5},\ref{eq:5'})
is a product of its irreps in both $\bm{k}$-space and in valley space.
In valley space, as shown in Table \ref{tab:irreps}, valley-triplet
order $\tau_{1}$ transforms as the $A_{1}$ irrep (i.e. the gap does
not change sign upon either a $C_{2z}$ or $C_{2x}$ rotation), whereas
valley-singlet order $i\tau_{2}$ transforms as the $B_{2}$ irrep
(i.e. the gap changes sign upon either a $C_{2z}$ or $C_{2x}$ rotation).

To determine the leading SC instabilities, one needs to solve the
linearized pairing gap equation 
\begin{align}
\hat{\Delta}(\omega,\bm{k})=\sum_{i,\omega',\bm{k}'} & D(\omega-\omega',\bm{k}-\bm{k}')\Lambda^{i}\hat{G}(\omega',\bm{k}')\nonumber \\
 & \times\hat{\Delta}(\omega',\bm{k}')\hat{G}^{T}(-\omega',-\bm{k}')\Lambda^{iT}\label{eq:gap}
\end{align}
where $\{\Lambda^{i}\}$ are the pertinent boson-fermion couplings
--- e.g., for inter-valley fluctuations we have $\{\Lambda^{i}\}=\{\tau_{1},\tau_{2}\}$.
Here, $D(\omega-\omega',\bm{k}-\bm{k}')$ is the bosonic propagator
multiplied by $g^{2}$ and renormalized by the boson-fermion coupling,
and  $\hat{G}(\omega,\bm{k})=\left[i\omega-\hat{\epsilon}(\bm{k})\right]^{-1}$, in matrix form,
is the fermionic Green's function. Because here we are only interested
in the momentum and spin/valley structure of the gap function, for
simplicity we take the kernel of the gap equation to be diagonal in
frequency and assume $\hat{\Delta}(\omega,\bm{k})=\hat{\Delta}(\bm{k})$.

\subsection{Inter-valley channel, $\phi_{v}$}

First, we focus on the pairing problem when the dominant fluctuations
are in the inter-valley channel, corresponding to either of the
two scalar bosonic fields that form $\phi_{v}=\left(\phi_{10},\phi_{20}\right)$
introduced in Sec.~\ref{sec:fluctuations}. In this case, we have
$\{\Lambda^{i}\}=\{\tau_{1}\sigma_{0},\tau_{2}\sigma_{0}\}$ in Eq.~\eqref{eq:gap}.
Before solving Eq.~\eqref{eq:gap}, it is instructive to convert
{the effective interaction} $-D(k-p)\sum_{i=1,2}\left[\Psi^{\dagger}(k)\tau_{i}\Psi(p)\right]\left[\Psi^{\dagger}(-k)\tau_{i}\Psi(-p)\right]$
to the particle-particle channel to determine for which pairing symmetry
the interaction is attractive. To this end, we use the following Fierz
identity for the particle-hole channel~\cite{Joern2018}
\begin{align}
\sum_{i=1,2}(\tau_{i})_{\alpha\beta}(\tau_{i})_{\mu\nu}=(\tau_{1})_{\alpha\mu}(\tau_{1}^T)_{\nu\beta}-(i\tau_{2})_{\alpha\mu}(i\tau_{2}^{T})_{\nu\beta}.\label{eq:fierz1}
\end{align}
Plugging it into the effective interaction, we find 
\begin{align}
 & -D(k-p)\sum_{i=1,2}\left[\Psi^{\dagger}(k)\tau_{i}\Psi(p)\left]\right[\Psi^{\dagger}(-k)\tau_{i}\Psi(-p)\right]\nonumber \\
 & ={-}D(k-p)\left[\Psi^{\dagger}(k)\tau_{1}\Psi^{\dagger T}(-k)\left]\right[\Psi^{T}(-p)\tau_{1}\Psi(p)\right]\nonumber \\
 & {+}D(k-p)\left[\Psi^{\dagger}(k)i\tau_{2}\Psi^{\dagger T}(-k)\left]\right[\Psi^{T}(-p)i\tau_{2}^{T}\Psi(p)\right].
\end{align}
%{(YW: I just discovered a stupid sign error going from Eq. 10 to Eq. 11. Other similar errors have also been corrected. Sorry for the confusion.)}
%Recall that, in the particle-particle channel, a positive coefficient corresponds to attraction. \textcolor{magenta}{\textbf{RMF: I think this should be explained, it won't be obvious to people that have not done this type of calculation before, which is probably the case of many people in the TBG community}} 
Therefore, the pairing interaction mediated
by inter-valley fluctuations is attractive in the valley-triplet channel
and repulsive in the valley-singlet channel, analogous to the case
of spin ferromagnetic fluctuations. However, as we mentioned above,
from a symmetry point of view the valley singlet and triplet channels
are always mixed.

To determine the composition of valley singlet and triplet components, we go back to Eq.~\eqref{eq:gap} and inserting the Fierz identity
\eqref{eq:fierz1}, and obtain
\begin{align}
 & \hat{\Delta}(\omega,\bm{k})=\frac{1}{2}\sum_{\omega,\bm{k}}D(\omega-\omega',\bm{k}-\bm{k}')\nonumber \\
 & \times\left\{ \tau_{1}\Tr\left[\tau_{1}\hat{G}(\omega',\bm{k}')\hat{\Delta}(\omega',\bm{k}')\hat{G}(-\omega',-\bm{k}')\right]\right.\nonumber \\
 & \left.-i\tau_{2}\Tr\left[i\tau_{2}^{T}\hat{G}(\omega',\bm{k}')\hat{\Delta}(\omega',\bm{k}')\hat{G}(-\omega',-\bm{k}')\right]\right\} \label{eq:gap2}
\end{align}
If our system had valley SU(2) symmetry, $\hat{G}$ would be proportional
to the identity matrix. In this case, the valley-triplet solution
with $\hat{\Delta}\propto\tau_{1}$ would be the eigenvector corresponding
to the leading pairing channel. The valley-singlet pairing channel,
$\hat{\Delta}\propto i\tau_{2}$, would only have the trivial solution
$\hat{\Delta}=0$ due to the minus sign in the second term inside
the brackets.

Our band dispersion, however, does not have valley SU(2) symmetry.
Thus, one needs to solve \eqref{eq:gap2} as a matrix equation and
find its eigenvectors, which in general will be a mixture of valley-singlet
and valley-triplet. Nonetheless, since we expect $\hat{\Delta}(\bm{k})$
to be peaked around the valley hot-spots {[}see Fig.~\ref{fig:1}(a){]},
where the fermions from the two valleys are degenerate, the problem
can be simplified. Let us focus on one valley hot spot, around which
we assume the pairing gap has a very weak dependence on $\bm{k}$.
From Eq. \eqref{eq:dispersion}, the Green's function can be written
as
\begin{widetext} 
\begin{align}
\hat{G}(\omega,\pm\bm{k})= & \begin{pmatrix}{i\omega-\epsilon_{K}(\bm{k})} & 0\\
0 & {i\omega-\epsilon_{K}'(\bm{k})}
\end{pmatrix}^{-1}= -\frac{i\omega+\epsilon_{+}\pm\epsilon_{-}\tau_{3}}{(i\omega-\epsilon_{+})^{2}-\epsilon_{-}^{2}}
\end{align}
where $\epsilon_{\pm}=\left[\epsilon_{K}(\bm{k})\pm\epsilon_{K'}(\bm{k})\right]/2$.
Near the hot-spots, we linearize the dispersion $\epsilon_{K}(\bm{k})=\bm{v}_{F}\cdot\bm{k}$,
such that the momentum integral becomes $c\int d\epsilon_{+}d\epsilon_{-}$,
where the constant $c$ is the Jacobian of the transformation. Within
this approximation, one can verify that the valley-triplet pairing solution
$\hat{\Delta}(\omega,\bm{k})=\Delta(\omega)\tau_{1}$ remains an eigenvector:
\begin{align}
\Delta(\omega)\tau_{1}= & {c}\sum_{\omega}\int d\epsilon_{+}d\epsilon_{-}\Delta(\omega')D(\omega-\omega',\bm{k}-\bm{k}')\left\{ \tau_{1}\Tr\left[\tau_{1}\frac{i\omega'+\epsilon_{+}+\epsilon_{-}\tau_{3}}{(i\omega-\epsilon_{+})^{2}-\epsilon_{-}^{2}}\tau_{1}\frac{-i\omega'+\epsilon_{+}-\epsilon_{-}\tau_{3}}{(-i\omega-\epsilon_{+})^{2}-\epsilon_{-}^{2}}\right]\right.\nonumber \\
 & \left.-i\tau_{2}\Tr\left[i\tau_{2}^{T}\frac{i\omega'+\epsilon_{+}+\epsilon_{-}\tau_{3}}{(i\omega-\epsilon_{+})^{2}-\epsilon_{-}^{2}}\tau_{1}\frac{-i\omega'+\epsilon_{+}-\epsilon_{-}\tau_{3}}{(-i\omega-\epsilon_{+})^{2}-\epsilon_{-}^{2}}\right]\right\} \nonumber \\
= & {2c}\sum_{\omega}\int d\epsilon_{+}d\epsilon_{-}\Delta(\omega')D(\omega-\omega',\bm{k}-\bm{k}')(\tau_{1})\left[\frac{\omega'^{2}+\epsilon_{+}^{2}+\epsilon_{-}^{2}}{\left(\omega^{2}+(\epsilon_{+}+\epsilon_{-})^{2}\right)\left(\omega^{2}+(\epsilon_{+}+\epsilon_{-})^{2}\right)}\right]\nonumber \\
= & 2c\sum_{\omega',\bm{k}'}{\Delta}(\omega')\tau_{1}\,D(\omega-\omega',\bm{k}-\bm{k}')\left[\frac{1}{\omega'^{2}+\left(\epsilon_{\bm{K}}(\bm{k})\right)^{2}}+\frac{1}{\omega'^{2}+\left(\epsilon_{\bm{K}'}(\bm{k})\right)^{2}}\right]\label{eq:gap3}
\end{align}
\end{widetext} Note that the $i\tau_{2}\Tr\left(\cdots\right)$ term
in the first line, which would have given a mixture with valley-singlet,
vanishes after the momentum integration, since the trace is odd in
$\epsilon_{-}$ ; this is approximately true when linearizing the dispersion.
The integration leads to the standard Cooper logarithmic instability
in the valley-triplet channel. The same calculation in the valley-singlet
channel yields only the trivial solution $\Delta=0$.

\begin{figure}[htp]
\subfigure[\label:]{\includegraphics[width=0.45\columnwidth]{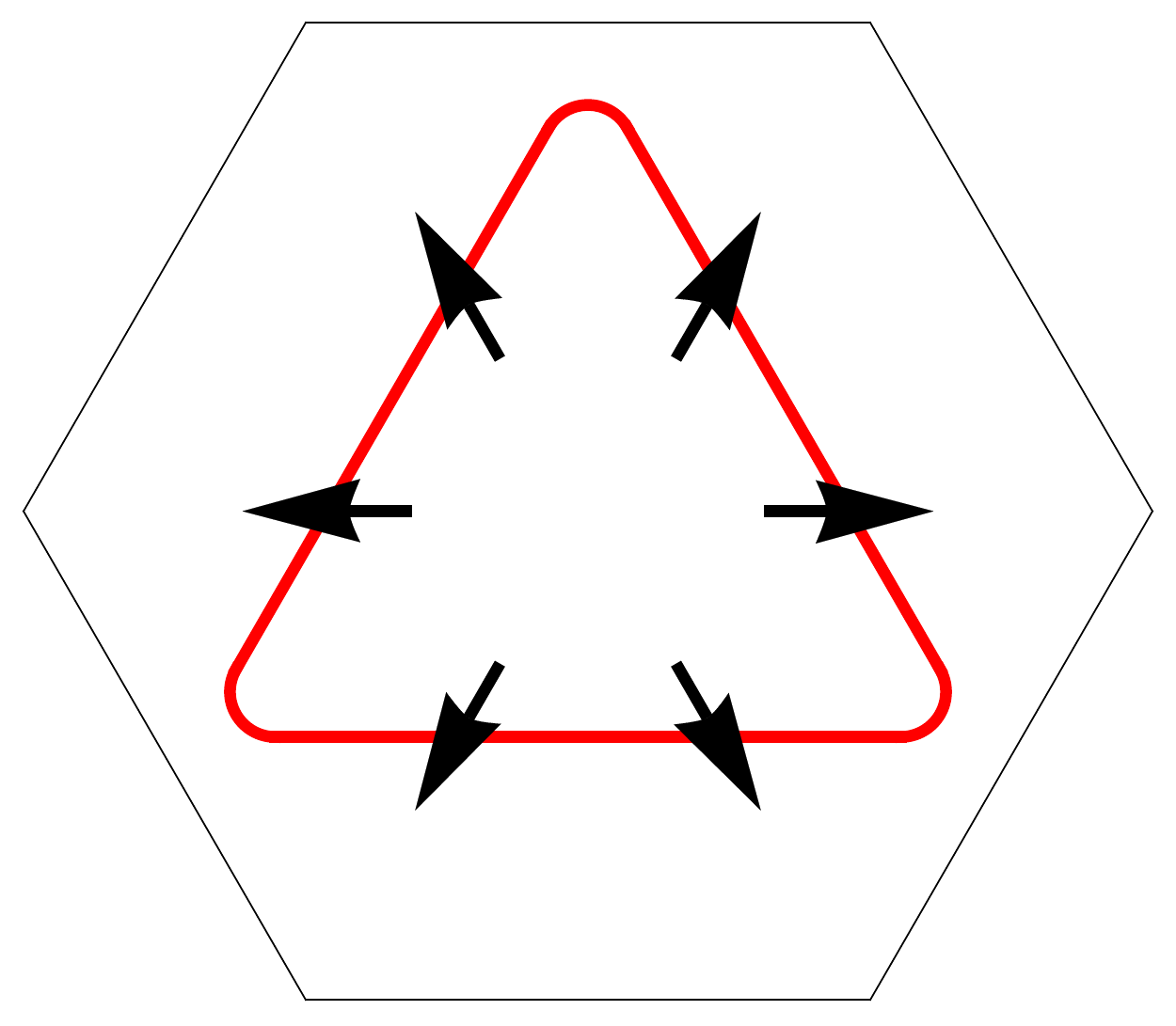}~~~\includegraphics[width=0.45\columnwidth]{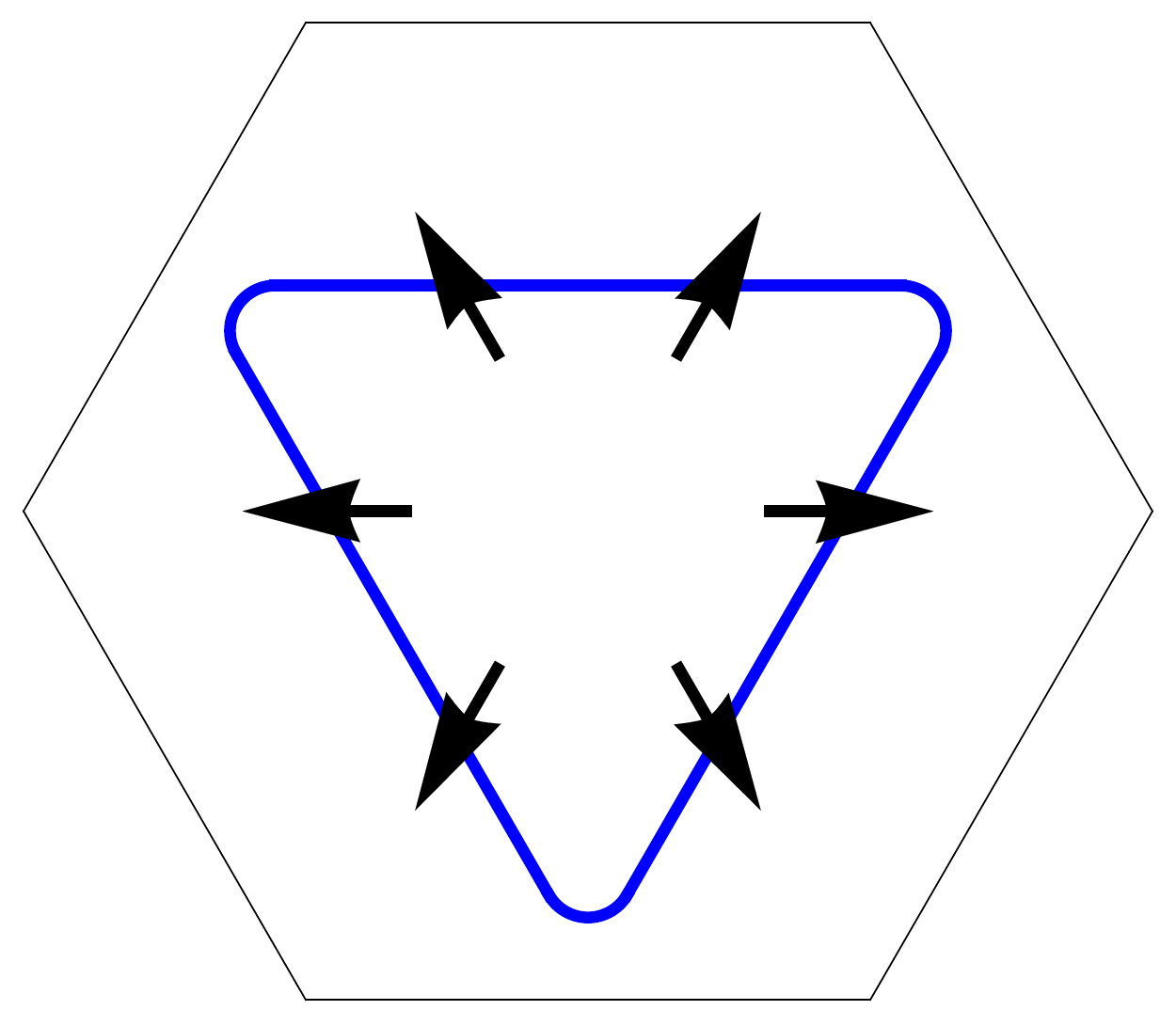}}
%	\subfigure[\label{Fig:v:E12}]{}
 \subfigure[\label:]{\includegraphics[width=0.45\columnwidth]{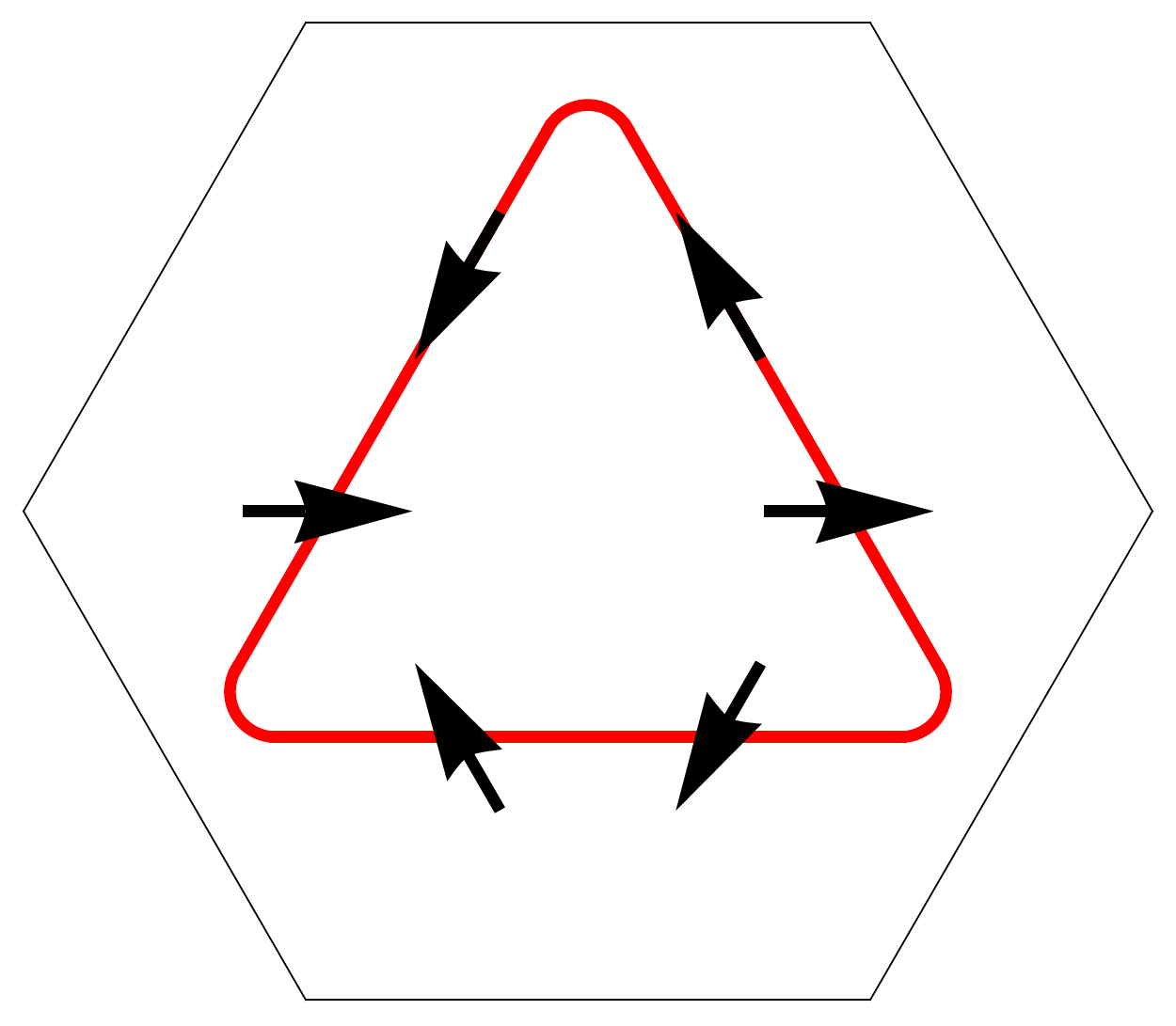}~~~\includegraphics[width=0.45\columnwidth]{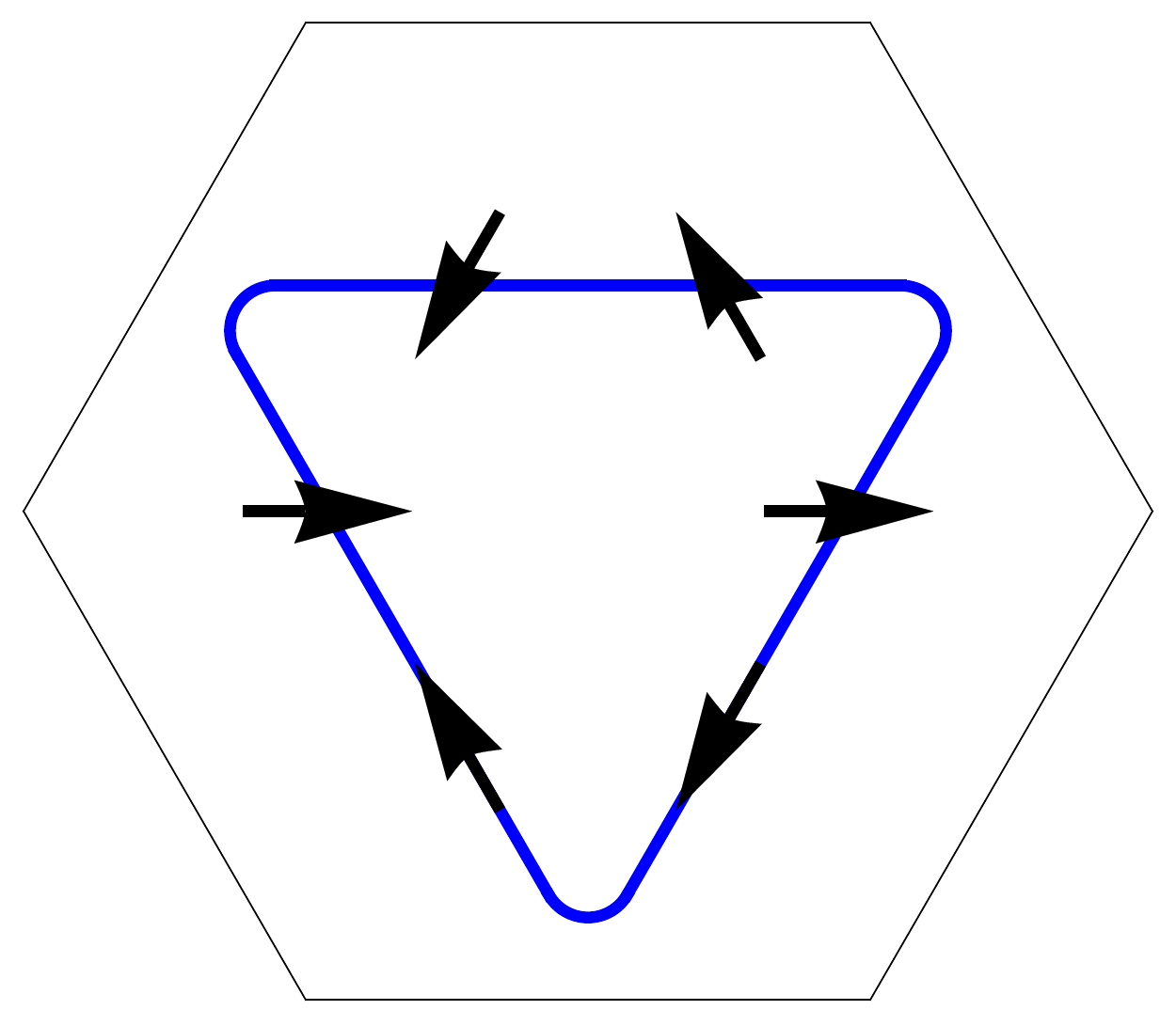}}
\caption{Illustration of the $E_{1}$ ($p\pm ip$) {[}panel (a){]} and $E_{2}$
($d\pm id$) {[}panel (b){]} valley-triplet pairing states mediated
by inter-valley flucutations.}
\label{Fig:PhivPairing} 
\end{figure}

Going back to Eqs.~\eqref{eq:5} and \eqref{eq:5'}, the full symmetry
of the pairing order parameter depends on its structure in the spin
sector. If it is spin-singlet, restricting the analysis to fully gapped
states only, it can be of $A_{1}$ ($s$-wave) or $E_{2}$ ($d$-wave)
symmetry; if it is spin-triplet, the fully gapped state corresponds
to a $E_{1}$ ($p$-wave) pairing. Because the valley-triplet component
of the gap, $\tau_{1}$, transforms trivially under $D_{6}$ (i.e.
as the $A_{1}$ irrep), the $\bm{k}$-space structure of the gap solely
determines the irrep of the order parameter. Energetically, it is
natural to expect that $\Delta_{o}\sim(\cos\theta,\pm i\sin\theta)$
for the $E_{1}$ state (yielding a $p\pm ip$ state) and $\Delta_{e}\sim(\cos2\theta,\pm i\sin2\theta)$
for the $E_{2}$ state (yielding a $d\pm id$ state) to ensure a fully
gapped state, where $\theta$ is a polar angle parametrizing the FS
(without loss of generality, in the ensuing discussion we will choose
the plus sign). We illustrate the $p+ip$ and $d+id$ SC gaps corresponding
to these $E_{1}$ and $E_{2}$ pairing states in Fig.~\ref{Fig:PhivPairing};
the arrows correspond to the direction of $\Delta_{o,e}$ in the two-dimensional
space of their corresponding irrep $E_{1,2}$. Since the winding on
the two valley FSs are the same, these are both chiral states
that also break time-reversal symmetry. From the winding numbers of
the gap functions on the FS, the Chern number of the $E_{2}$ state
is $C=4$ (using the convention that a spinless $p+ip$ superconductor has Chern number $C=1/2$ due to the redundancy of the Bogoliubov-de-Gennes Hamiltonian~\cite{liu-2018}), whereas for the $E_{1}$ state
it is $C=2$. On the other hand, the $A_{1}$ state has a Chern number
$C=0$. Note that topological SC in TBG was previously proposed within
different models \cite{Leon1}.

Note that the inter-valley fluctuations do not directly distinguish
the strengths of spin-singlet and spin-triplet pairing tendencies.
Moreover, the pairing interaction is peaked at small momentum transfer,
which is known to favor multiple nearly degenerate pairing orders
\cite{Kivelson,Gastiasoro20}. We thus expect all three pairing instabilities
to be close competitors. Which order is the true leading instability
requires additional details beyond the scope of our model, such as
the correlation length of the nematic fluctuations and the fermionic
band dispersion. Extrinsic factors such as disorder also could also
lift this near-degeneracy, likely favoring the time-reversal symmetric
$A_{1}$ state.

\subsection{Spin-valley coupled channel, $\phi_{sv}$}

We now consider spin-valley coupled fluctuations, which are associated with either
of the two vector bosonic fields $\phi_{sv}=\left(\phi_{1i\neq0},\phi_{2i\neq0}\right)$
(see Sec. \ref{sec:fluctuations}). The pairing gap equation is
still given by Eq.~\eqref{eq:gap}, but now with $\{\Lambda^{i}\}=\{\bm{\sigma}\tau_{1},\bm{\sigma}\tau_{2}\}$.
Just like in the $\phi_{v}$ case, although the gap function is a
mixture between valley-singlet and valley-triplet, near the valley
hot-spots the valley singlet/triplet assignment remains approximately
valid. To determine the leading attractive pairing, we start from
the Fierz identities: 
\begin{align}
 & \sum_{i=1,2}(\tau_{i})_{\alpha\beta}(\tau_{i})_{\mu\nu}=(\tau_{1})_{\alpha\mu}(\tau_{1})_{\nu\beta}-(i\tau_{2})_{\alpha\mu}(i\tau_{2}^{T})_{\nu\beta}.\nonumber \\
 & (\bm{\sigma})_{\gamma\delta}\cdot(\bm{\sigma})_{\rho\lambda}=\frac{1}{2}(\bm{\sigma}i\sigma_{2})_{\gamma\rho}\cdot(i\sigma_{2}^{T}\bm{\sigma}^{T})_{\lambda\delta}-\frac{3}{2}(i\sigma_{2})_{\gamma\rho}(i\sigma_{2}^{T})_{\lambda\delta}\label{eq:fierz2}
\end{align}

Taking the direct product, we find: 
\begin{align}
 & -D(k-p)\sum_{i=1,2}\left[\Psi^{\dagger}(k)\tau_{i}\bm{\sigma}\Psi(p)\left]\cdot\right[\Psi^{\dagger}(-k)\tau_{i}\bm{\sigma}\Psi(-p)\right]\nonumber \\
= &-\frac{3}{2}D(k-p)\left[\Psi^{\dagger}(k)i\sigma_{2}i\tau_{2}\Psi^{\dagger T}(-k)\left]\right[\Psi(p)i\tau_{2}^{T}i\sigma_{2}^{T}\Psi^{T}(-p)\right]\nonumber \\
 & -\frac{1}{2}D(k-p)\left[\Psi^{\dagger}(k)(i\bm{\sigma}\sigma_{2})(\tau_{1})\Psi^{\dagger T}(-k)\right]\nonumber \\
 & ~~~~~~~~\times\left[\Psi(p)(\tau_{1})(i\sigma_{2}^{T}\bm{\sigma}^{T})\Psi^{T}(-p)\right]\nonumber \\
 & +\frac{3}{2}D(k-p)\left[\Psi^{\dagger}(k)i\sigma_{2}(\tau_{1})\Psi^{\dagger T}(-k)\right]\nonumber \\
 & ~~~~~~~~\times\left[\Psi(p)(\tau_{1})i\sigma_{2}^{T}\Psi^{T}(-p)\right]\nonumber \\
 & +\frac{1}{2}D(k-p)\left[\Psi^{\dagger}(k)(i\bm{\sigma}\sigma_{2})\tau_{3}\Psi^{\dagger T}(-k)\right]\nonumber \\
 & ~~~~~~~~\times\left[\Psi(p)\tau_{3}(i\sigma_{2}^{T}\bm{\sigma}^{T})\Psi^{T}(-p)\right]\label{eq:14}
\end{align}

Therefore, the most attractive (i.e.~negative in the above equation)
interaction is in the spin-singlet, valley-singlet channel. Note that
there is also sub-leading attraction in one spin-triplet valley-triplet
channel, but repulsion in the spin-singlet valley-triplet channel
and in another spin-triplet valley-triplet channel.

\begin{figure}[htp]
\subfigure[\label:]{\includegraphics[width=0.45\columnwidth]{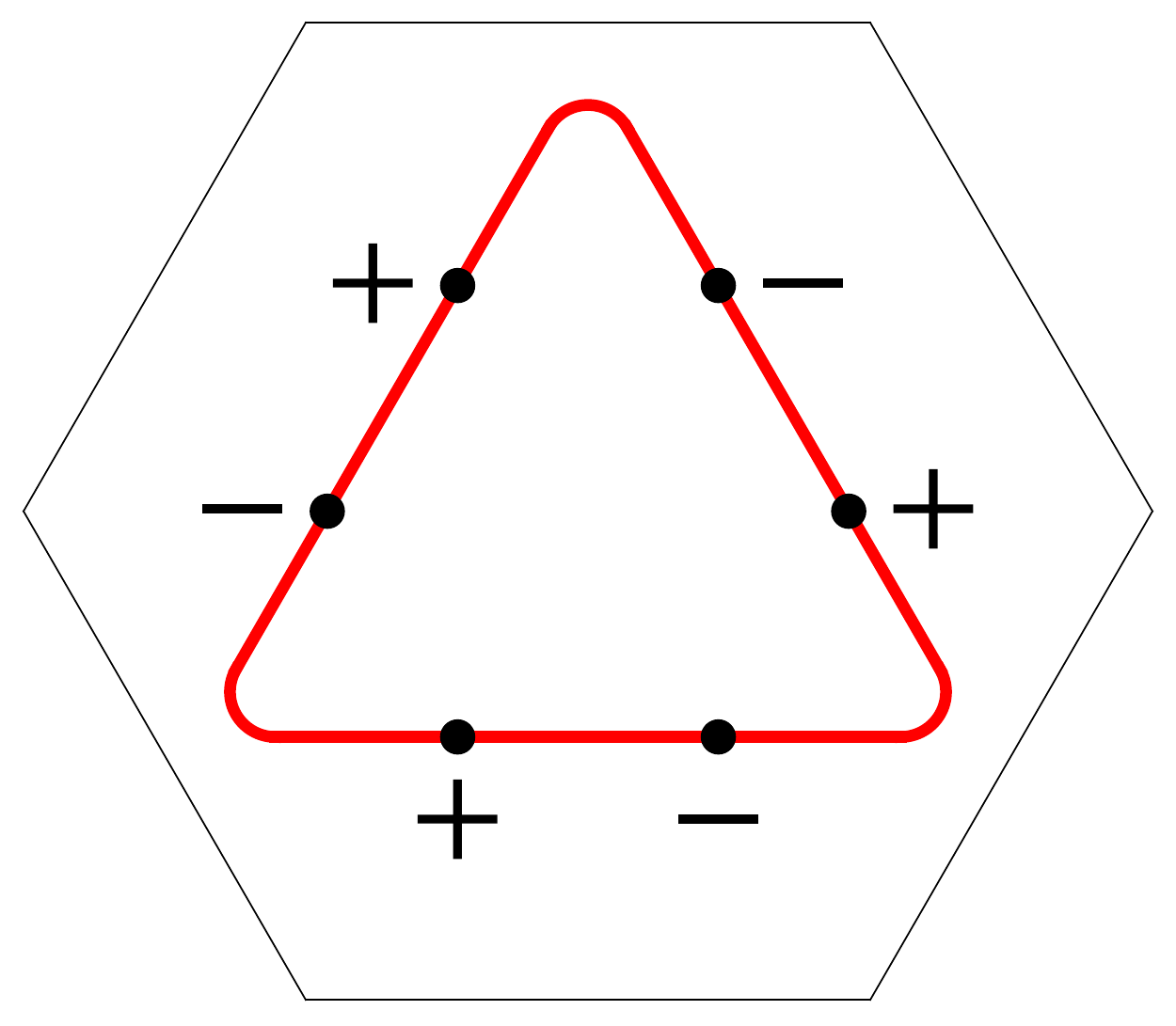}~~~\includegraphics[width=0.45\columnwidth]{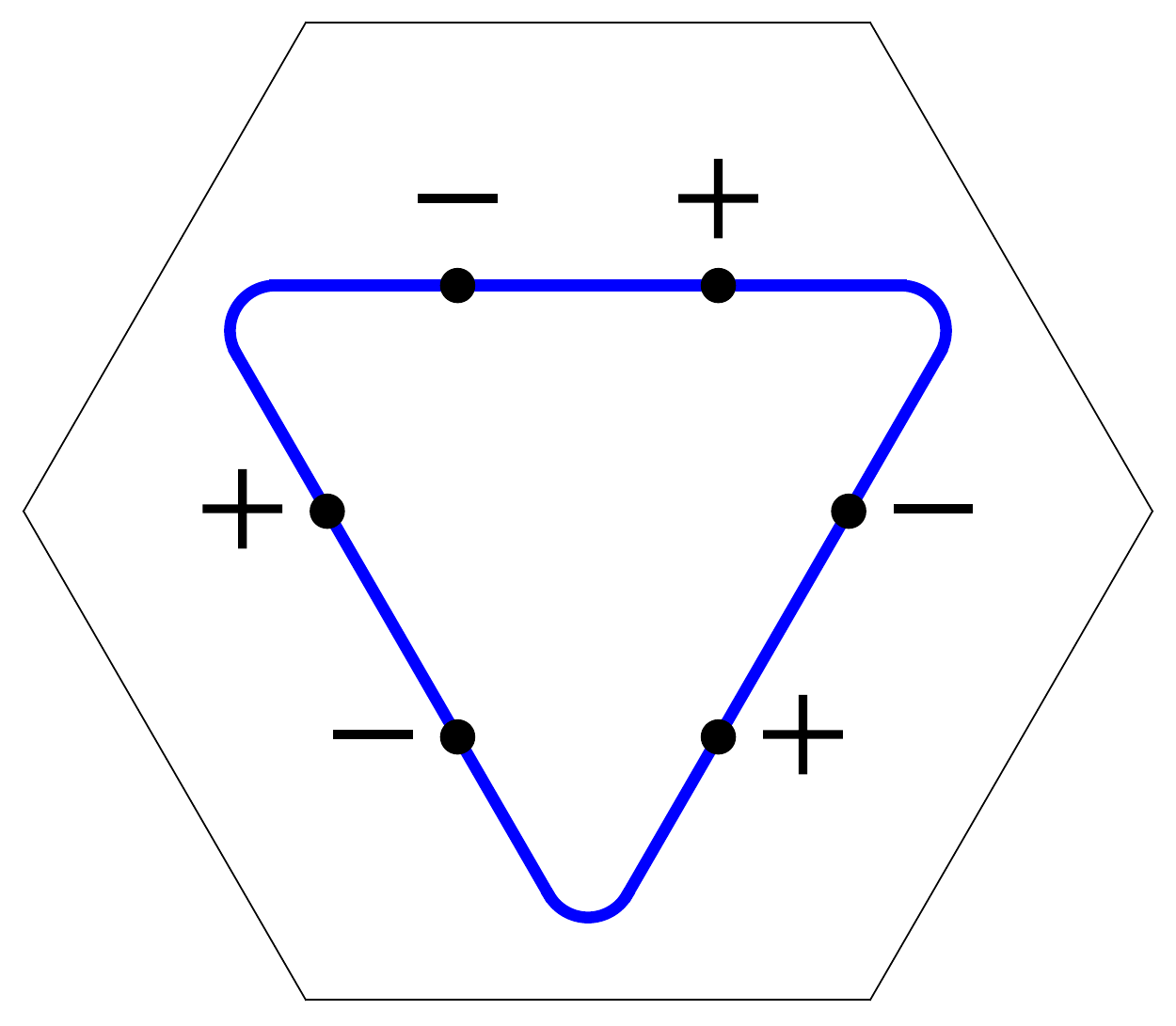}}
\subfigure[\label:]{\includegraphics[width=0.45\columnwidth]{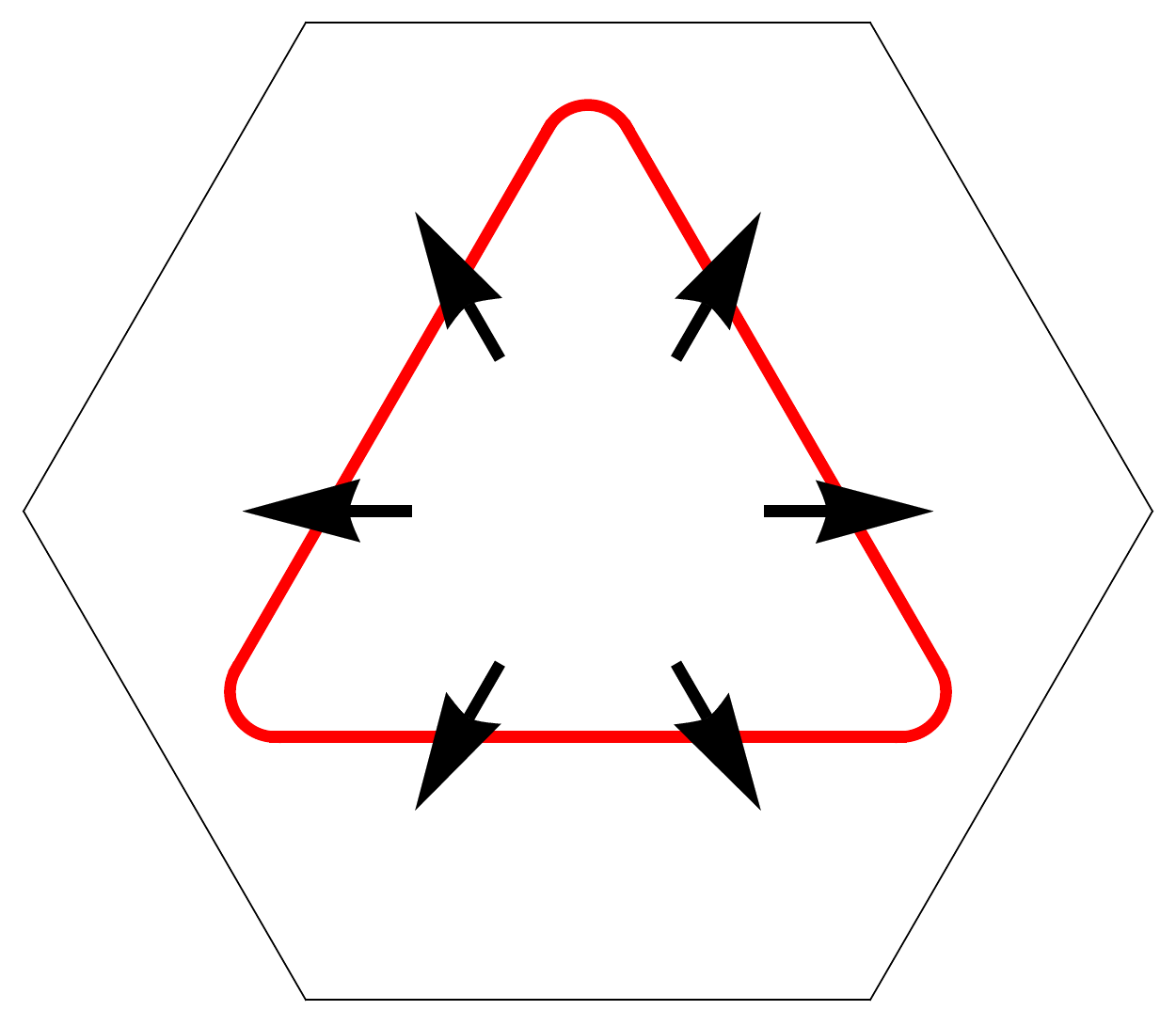}~~~\includegraphics[width=0.45\columnwidth]{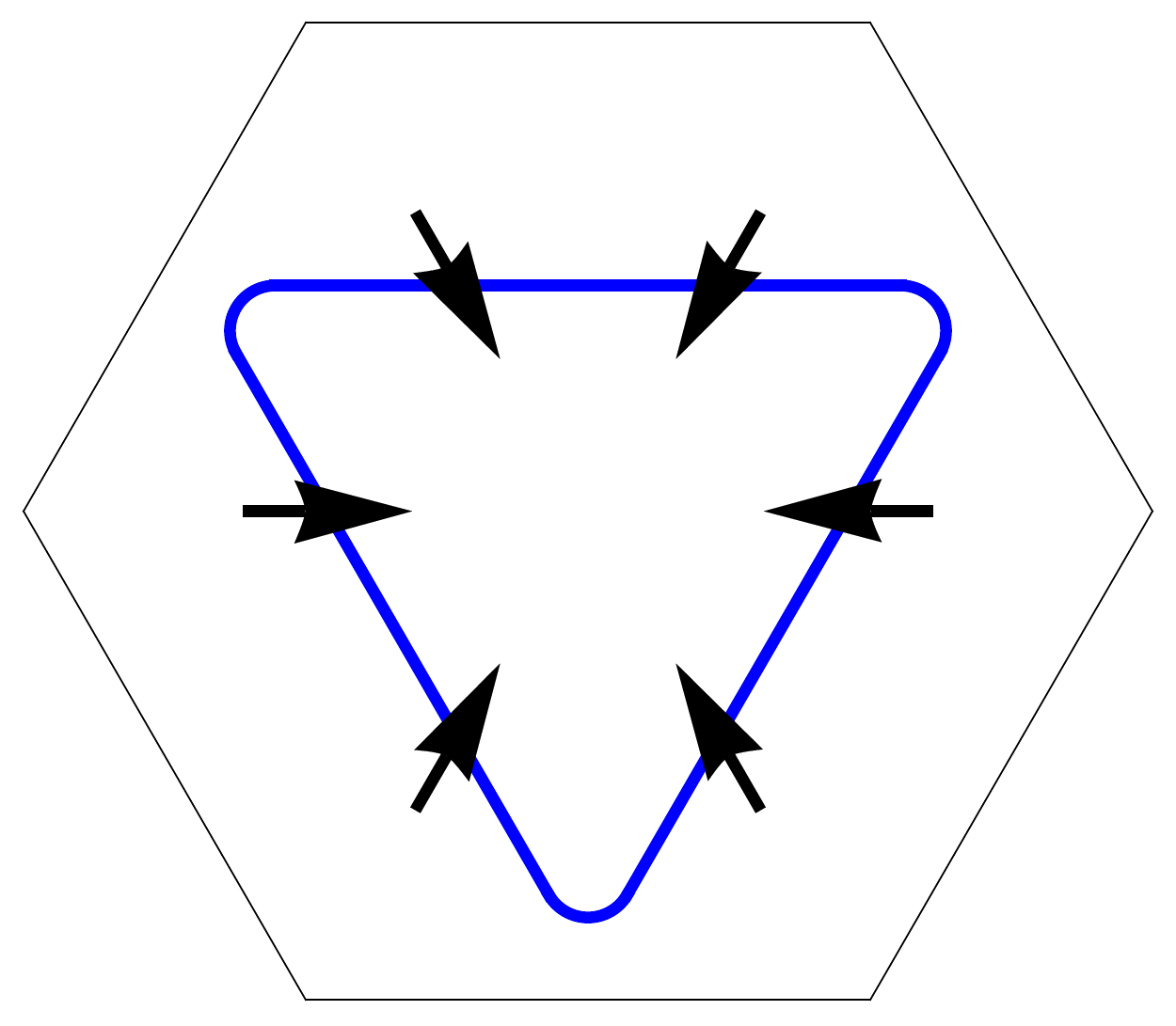}}
\caption{Illustration of the $A_{2}$ ($i$-wave) {[}panel (a){]} and $E_{2}$
($d\pm id$) {[}panel (b){]} valley-singlet pairing states mediated
by spin-valley flucutations.}
\label{Fig:PhisvPairing} 
\end{figure}

By solving the gap equation with a valley-split Green's function like
we did in Eq.~\eqref{eq:gap3}, we confirm that indeed the leading
pairing instability is towards a spin-singlet, valley-singlet channel.
From Eq.~\eqref{eq:5}, we see that this is an even-parity superconducting
gap with a dominant odd spatial function $\Delta_{o}(\bm{k})$. There
are thus three choices, $\Delta_{o}\sim\cos\theta,\sin\theta,\cos3\theta$;
note that the gap function $\sin3\theta$ vanishes on the valley hot
spots, and thus can be discarded. The first two choices for $\Delta_{o}$
give rise to a gap function that transforms as the two-dimensional
irrep $E_{2}$ (i.e.~$d$-wave); the resulting superconducting state
is fully gapped if the chiral  ``$d+id$" solution is realized, $\Delta_{o}\sim\left(\cos\theta,\,i\sin\theta\right)$.
To see that the total gap function transforms as $E_{2}$, note that
$\Delta_{o}\sim\left(\cos\theta,\,i\sin\theta\right)$ transforms
as the $E_{1}$ irrep, and that the valley component $i\tau_{2}$
transforms as the $B_{2}$ irrep. The irrep associated with the full
gap is thus obtained by combining the valley and spatial components
using the product $E_{1}\otimes B_{2}=E_{2}$. If, on the other hand, $\Delta_{o}\sim\cos3\theta$ is selected,
the total gap function transforms as the irrep $A_{2}$ ($i$-wave).
This follows from the fact that $\cos3\theta$ transforms as the $B_{1}$
irrep and that $B_{1}\otimes B_{2}=A_{2}$. Note that a similar $i$-wave
gap was previously proposed in another study on TBG \cite{Chichinadze19}.
While a single-band $i$-wave order has 12 nodes, our $A_{2}$ order
parameter has six nodes per valley FS, located between the valley
hot-spots. We illustrate the pairing orders $A_{2}$ and $E_{2}$
in Fig.~\ref{Fig:PhisvPairing}. Note that for $E_2$ in Fig.~\ref{Fig:PhisvPairing}(b), the windings on the two FS's are the same, thus breaking time-reversal symmetry. Although it has the
same symmetry as a $d+id$ state, the winding number of $\Delta_{o}\sim\left(\cos\theta,\,i\sin\theta\right)$
is one, suggesting that the total Chern number is $C=2$ (including
two valleys). The gapless $A_{2}$ pairing state
does not have a well-defined Chern number due to the existence of
nodes. As before, to further select between the two pairing instabilities,
additional microscopic details must be considered. For example, disorder
effects are expected to favor the fully gapped $E_{2}$ state over
the nodal $A_{2}$ state. %\textcolor{magenta}{\textbf{RMF What is the meaning of gaps of opposite signs at the same valley hot spot (like in the i-wave case)? Does it really imply nodes?}} {YW: The node is between valley hot spots (three corners of the FS in Fig. 4(a)). No nodes at the hot spots.}

\subsection{Ferromagnetic channel, $\phi_{s}$}

In the case of enhanced ferromagnetic fluctuations $\phi_{s}$ near
a van Hove singularity, the linearized gap equation is given by \eqref{eq:gap}
with $\{\Lambda_{i}\}=\{\bm{\sigma}\}$. Using the second line of
Eq.~\eqref{eq:fierz2} we find that the leading instability is in
the spin-triplet channel $\hat{\Delta}\propto\bm{\sigma}i\sigma_{2}$
and thus the favored pairing channel is of odd-parity. The relevant
odd parity irreps of $D_{6}$ are $E_{1}$ ($p$-wave) and $B_{1,2}$
($f$-wave).

\begin{figure}[htp]
\includegraphics[width=0.4\columnwidth]{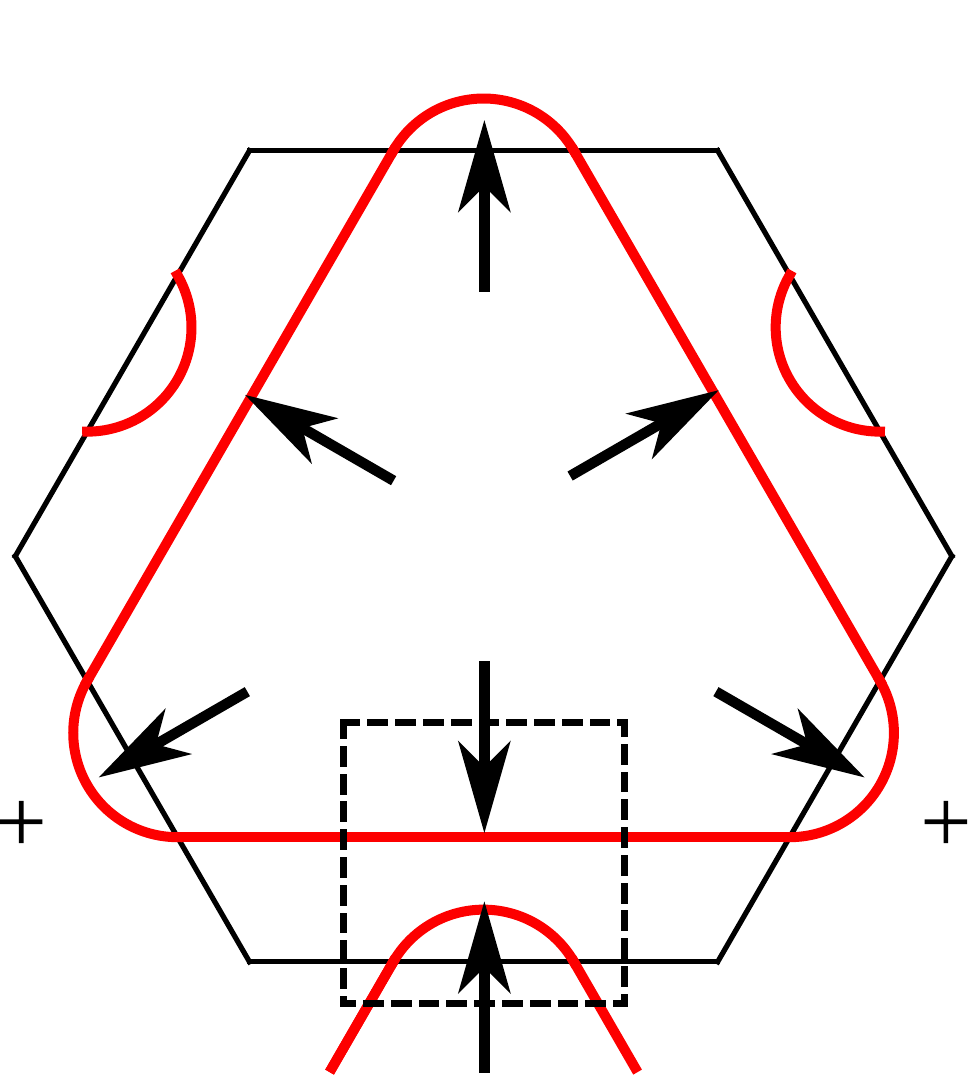}
\caption{The structure of $\Delta'_{o}$ for the case of valley-triplet, spin-triplet $p + i p$ pairing mediated by ferromagnetic fluctuations. Note the sign change between a pair of van Hove hot-spots. } 
%\textcolor{magenta}{\textbf{RMF I think this figure is confusing, because the gap is a vector, not a scalar. Could we redo it?}}}
\label{fig:scfs}
\end{figure}

\begin{figure}[htp]
\subfigure[\label:]{\includegraphics[width=0.9\columnwidth]{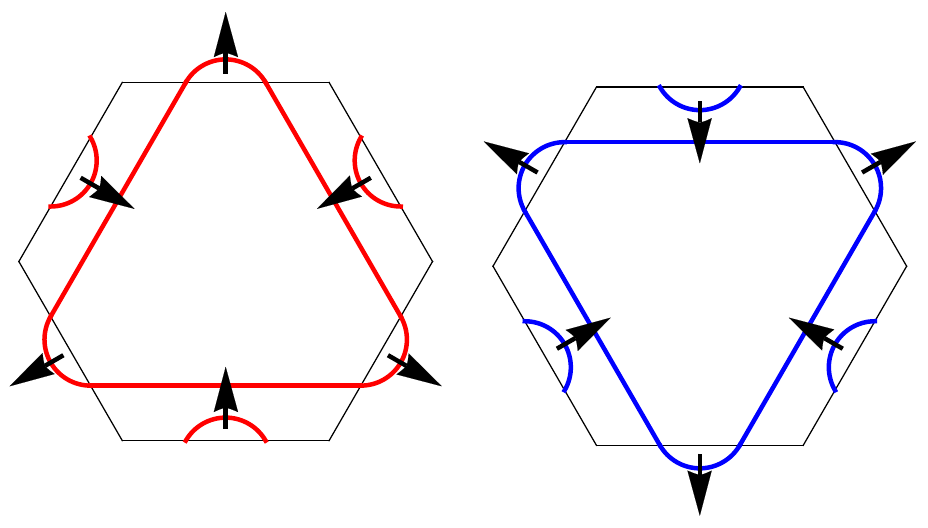}}
\subfigure[\label:]{\includegraphics[width=1\columnwidth]{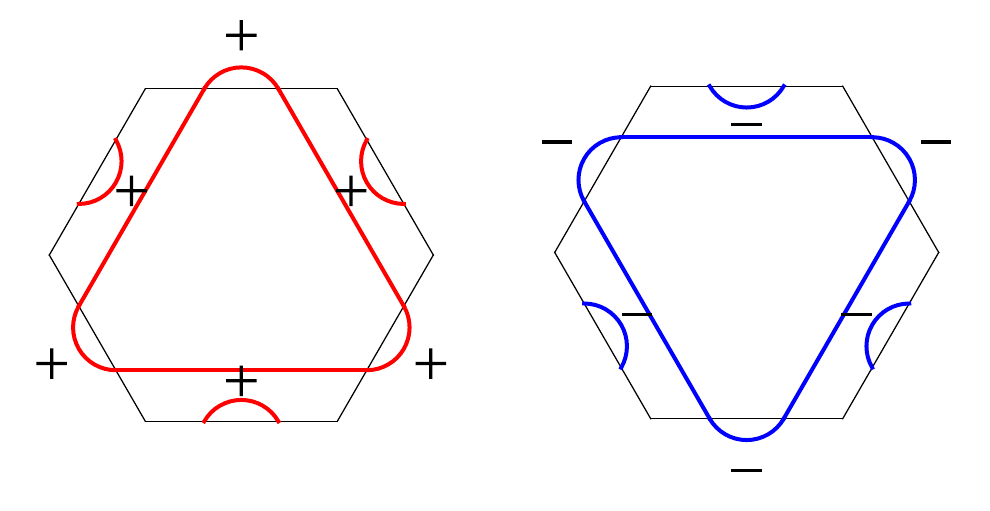}}
\caption{Illustration of the   $E_{1}$
($p\pm ip$) {[}panel (a){]} and $B_{2}$ ($f$-wave) {[}panel (b){]} spin-triplet pairing states mediated
by the ferromagnetic flucutations.}
\label{Fig:PhisPairing} 
\end{figure}

The $E_{1}$ irrep is two-dimensional, and depending on the gap structure
in the valley sector, there are two choices for the $\bm{k}$ dependence
of the gap function that are fully gapped ($p\pm ip$ state): $\Delta_{o}'\sim\left(\cos\theta,\,i\sin\theta\right)$,
and $\Delta_{e}'\sim\left(\cos2\theta,\,i\sin2\theta\right)$. To
see this, we notice that $\Delta_{o}'$ transforms as $E_{1}$, and
the triplet valley component $\tau_{1}$ transforms as $A_{1}$; on
the other hand, $\Delta'_{e}$ transforms as $E_{2}$, and the valley-singlet
component $i\tau_{2}$ transforms as $B_{2}$, resulting in a $E_{1}$
total gap since $E_{2}\otimes B_{2}=E_{1}$. Because $\Delta'_{o}$
has opposite signs on the two van Hove hot-spots {[}see Fig.~\ref{fig:scfs}{]},
this configuration is expected to be energetically disfavored and
thus subleading to $\Delta'_{e}$, which is shown in Fig.~\ref{Fig:PhisPairing}(a).
However, because $\Delta_{e}'$ and $\Delta_{o}'$ result in a gap
with the same $E_{1}$ symmetry, they are always mixed. Since their
winding number on each FS is different, the Chern number of the resulting
SC order depends on whether $\Delta'_{o}$ or $\Delta'_{e}$ dominates.
For our case $|\Delta'_{e}|>|\Delta'_{o}|$, in each triangular FS
the phase of the order parameter winds \emph{twice}, like a $d+id$
order, despite the fact that the symmetry of the pairing state is
identical to $p+ip$. As a result, the total Chern number should be
$4$, coming from $C=2$ for two spin species. 

As for the case of a gap transforming as either the $B_{1}$ or $B_{2}$
irrep, it is possible to construct a fully gapped valley-singlet $B_{2}$
state with $\Delta'_{e}\sim\textrm{const}$, since $i\tau_{2}$ transforms
as $B_{2}$. For energetic reasons, we expect this gap to be favored
over the other possible combinations. The full gaps on the two valley
FSs have opposite signs, since this is a valley-singlet state, as
shown in Fig. \eqref{Fig:PhisPairing}(b). Superimposing the two Fermi
surfaces, the gap indeed has the $\sin3\theta$ variation across the
Brillouin zone characteristic of a $B_{2}$ $f$-wave gap, yet the
state is fully gapped. This pairing state has a Chern number $C=0$.

It is interesting to consider how disorder effects distinguish the
two pairing states. Assuming each graphene sheet is atomically clean,
disorder in TBG occurs at the the moir\'e length scale, for example,
from a variation of the twist angle. Scattering by impurities at the
moir\'e length scale has small (atomic lattice) momentum transfer and
preserves valley index. Since the $B_{2}$ pairing gap does not change
sign within a valley, we expect it to be more robust than $E_{1}$
order against such long distance disorder.

\subsection{Spin-current channel, $\phi_{sc}$}

\begin{figure}[htp]
\subfigure[\label:]{\includegraphics[width=0.9\columnwidth]{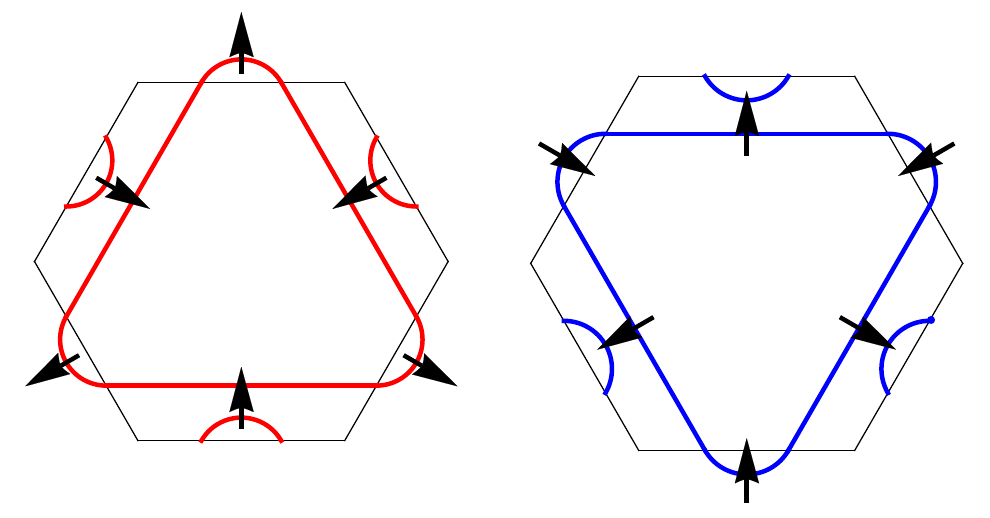}}
\subfigure[\label:]{\includegraphics[width=1\columnwidth]{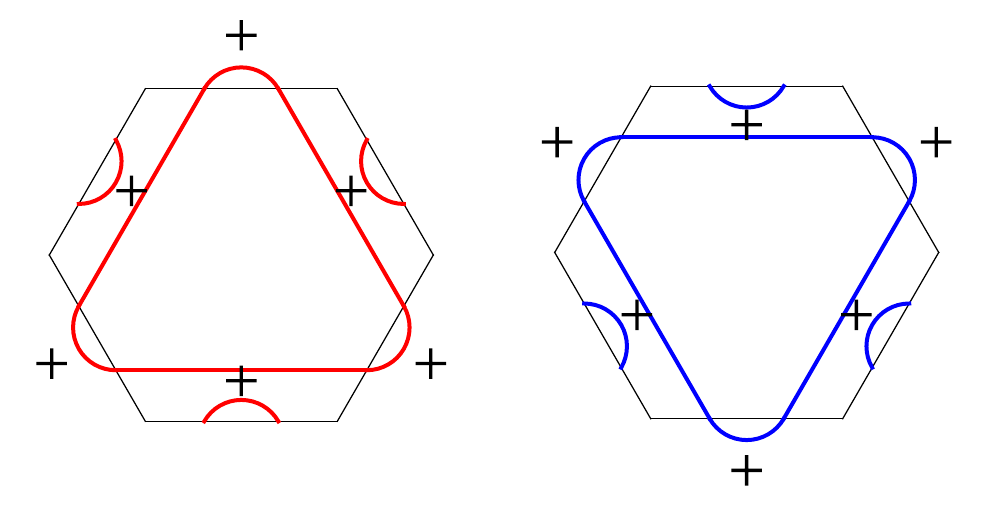}}
\caption{Illustration of  $E_{2}$
($d\pm id$) {[}panel (a){]}  and  the $A_{1}$ ($s$-wave) {[}panel (b){]} spin-singlet pairing states  mediated by
spin-current fluctuations.}
\label{Fig:PhiscPairing} 
\end{figure}

We now consider the fluctuations associated with the spin-current
state, described by the vector bosonic field $\phi_{sc}=\phi_{3i\neq0}$.
As we discussed, this bosonic field couples to the fermions via $\Lambda_{i}=\tau_{3}\sigma_{i}$.
Performing a similar calculation to Eq.~\eqref{eq:15}, we find that
the leading pairing instability remains in the spin-singlet channel.
This follows from the Fierz identities
\begin{align}
(\tau_{3})_{\alpha\beta}(\tau_{3})_{\mu\nu}= & \frac{1}{2}(\tau_{3})_{\alpha\mu}(\tau_{3})_{\nu\beta}+\frac{1}{2}(\tau_{0})_{\alpha\mu}(\tau_{0})_{\nu\beta}\label{eq:fierz3}\\
 & -\frac{1}{2}(\tau_{1})_{\alpha\mu}(\tau_{1})_{\nu\beta}-\frac{1}{2}(i\tau_{2})_{\alpha\mu}(i\tau_{2}^{T})_{\nu\beta}\nonumber \\
(\bm{\sigma})_{\gamma\delta}\cdot(\bm{\sigma})_{\rho\lambda}= & \frac{1}{2}(\bm{\sigma}i\sigma_{2})_{\gamma\rho}\cdot(i\sigma_{2}^{T}\bm{\sigma}^{T})_{\lambda\delta}-\frac{3}{2}(i\sigma_{2})_{\gamma\rho}(i\sigma_{2}^{T})_{\lambda\delta}\nonumber 
\end{align}
Taking a direct product, the fermion interaction can be rewritten
as 
\begin{align}
 & -D(k-p)\left[\Psi^{\dagger}(k)\tau_{3}\bm{\sigma}\Psi(p)\left]\cdot\right[\Psi^{\dagger}(-k)\tau_{3}\bm{\sigma}\Psi(-p)\right]=\nonumber \\
& -\frac{3}{4}D(k-p)\left[\Psi^{\dagger}(k)i\sigma_{2}\tau_{1}\Psi^{\dagger T}(-k)\left]\right[\Psi(p)\tau_{1}^{T}i\sigma_{2}^{T}\Psi^{T}(-p)\right]\nonumber \\
 &- \frac{3}{4}D(k-p)\left[\Psi^{\dagger}(k)i\sigma_{2}i\tau_{2}\Psi^{\dagger T}(-k)\left]\right[\Psi(p)i\tau_{2}^{T}i\sigma_{2}^{T}\Psi^{T}(-p)\right]\nonumber \\ 
 &+ \cdots\label{eq:15}
\end{align}
where $\cdots$ stand for terms with less negative (including positive)
coefficients corresponding to channels with weaker attraction or repulsion.
It is clear that the leading attractive channel for pairing is the
spin-singlet channel. From Eq.~\eqref{eq:5}, this corresponds to
an even-parity state. As before, the $\Delta_{e}\tau_{1}$ and $\Delta_{o}i\tau_{2}$
components are always mixed. Now, the valley-singlet pairing $i\tau_{2}$,
which transforms as $B_{2}$, requires $\Delta_{o}\sim\left(\cos\theta,\,i\sin\theta\right)$,
which transforms as $E_{1}$, resulting in the even-parity gap $E_{2}$.
However, $\Delta_{o}$ changes sign within a pair of van Hove hot-spots (see Fig.~\ref{fig:scfs}).
Since pairing is driven by the pairs of van Hove hot-spots, we conclude
that the energetically dominant component is valley triplet and spin
singlet ($\Delta_{e}\tau_{1}$), mixed with a small valley singlet
component. Since $\tau_{1}$ transforms as $A_{1}$, $\Delta_{e}$
can either be a constant, resulting in an $s$-wave gap, or $\Delta_{o}\sim\left(\cos2\theta,\,i\sin2\theta\right)$,
resulting in a $d$-wave gap. We illustrate these two pairing states
on the FSs in Fig.~\ref{Fig:PhiscPairing}. Compared with those promoted by
ferromagnetic fluctuations, the pairing gaps driven by spin-current
fluctuations have similar structures within an individual FS but differ
by parity. From their windings on the FS, the Chern numbers for the
$s$-wave state and $d$-wave state are $C=0$ and $C=4$, respectively.
At the hot-spots level, the $s$-wave and $d$-wave orders are degenerate.
However, when the full FS and/or disorder effects are taken into account,
it is likely that the fully gapped $s$-wave order becomes the dominant
one.

{\subsection{Valley-polarized channel, $\phi_{vp}$}}
Finally, we analyze the pairing instabilities driven by fluctuations in the valley-polarized channel. From the first line of Eq.~\eqref{eq:fierz3}, we see that the fluctuations of the $\phi_{vp}$ field yield attraction in two valley-triplet channels, with pairing orders $\sim \Delta\tau_0$ and $\sim\Delta\tau_3$. However, both of them correspond to intra-valley pairing. As we mentioned in the beginning of this Section, intra-valley pairing involves FSs that are not invariant under $C_{2z}$, i.e. the $\bm{k}$ and $\bm{-k}$ states are not at the same energy. As a result, the Cooper logarithm is cutoff and does not diverge at zero temperature. Hence, intra-valley pairing is energetically disfavored within the SU(4) spin-valley-fermion model. Therefore $\phi_{vp}$ fluctuations do not induce any superconducting phases, at least within weak-coupling. %\textcolor{magenta}{\textbf{RMF I added one sentence here.}}
\vspace{8mm}

\section{Nematic superconductivity}
\label{sec:nem}
Recent experiments in TBG have reported that the
SC dome near half-filling breaks the $C_{3z}$ rotational symmetry
of the moir\'e superlattice, i.e., it is a nematic superconductor \cite{Cao20,SigristUeda91}.
An interesting question is whether the pairing states that result
from the exhange of ferro-SU(4) fluctuations can become nematic. From
a symmetry standpoint, the nematic order parameter is a 3-state Potts
order parameter that transforms as the $E_{2}$ irrep \cite{Fernandes2}.
Thus, for a superconducting state to display subsidiary nematic order,
a SC bilinear must transform according to $E_{2}$. For an isolated
pairing instability, this is only possible for the $E_{1}$ ($p$-wave)
and $E_{2}$ ($d$-wave) instabilities, since $E_{1,2}\otimes E_{1,2}=A_{1}\oplus A_{2}\oplus E_{2}$.
However, as we discussed above, from energetic arguments one generally
expects that the SC ground state associated with the $E_{1,2}$ instabilities
will be the chiral one ($p\pm ip$ and $d\pm id$ states), since it
fully gaps the Fermi surface. In contrast, the nematic solutions
$p\pm p$ and $d\pm d$ generally generate nodes. This is consistent
with microscopic calculations discussed elsewhere, which argued that
the chiral solution is generally favored over the nematic one \cite{Kozii19}.
Nevertheless, it has been pointed out that proximity to a separate
normal-state nematic instability may tip the balance in favor of the
nematic over the chiral state \cite{Kozii19}.

Another possibility, which we explore in more details in this section,
is that superconducting nematic order can appear when two SC states
with different symmetries coexist microscopically. This was previously
discussed in the context of the iron pnictide superconductors, where
coexistence of $s$-wave and $d$-wave superconductivity gives rise
to a nematic SC state \cite{FernandesMillis}, and also in the context
of TBG, where nematic order was shown to arise from the coexistence of $i$-wave
and $d$-wave \cite{Chichinadze19}. As we discussed in Sec. \ref{sec:fluctuations},
the exchange of ferro-SU(4) fluctuations generally leads to closely
competing SC instabilities with different symmetries, which can potentially
lead to coexistence between different pairing states. Symmetry restricts
the possible combinations of gap functions that can result in a subsidiary
nematic order parameter. Quite generally, one needs to combine a state
that transforms as a one-dimensional irrep (i.e. $A_{1}$, $A_{2}$,
$B_{1}$, $B_{2}$) with another state that transforms as a two-dimensional
irrep (i.e. $E_{1}$, $E_{2}$), such that the product of these irreps
transforms as $E_{2}$. There are then four possibilities: $A_{1}$
and $E_{2}$ (dubbed $s\pm d$ state), $A_{2}$ and $E_{2}$ ($i\pm d$
state, as in Ref. \cite{Chichinadze19}), $B_{1}$ and $E_{1}$ ($f'\pm p$
state), $B_{2}$ and $E_{1}$ ($f\pm p$ state). The fact that these
states are nematic follows from the products of irreps of $D_{6}$:
$A_{1}\otimes E_{2}=A_{2}\otimes E_{2}=B_{1}\otimes E_{1}=A_{2}\otimes E_{2}=E_{2}$.
As shown in Table \ref{tab:summary}, we find that for each of the
ferro-SU(4) channels considered here that induce superconductivity, there is at least one combination
of closely competing SC states of different symmetries that can result
in an accompanying nematic order. As we mentioned, when two SC order
parameters $\Delta_{a,b}$ coexist, the energetically-favored coexisting
order parameter is typically of the form $\Delta_{a}\pm i\Delta_{b}$, breaking
time-reversal symmetry rather than spatial symmetries. However, as
we show below, owing to the internal structure of one of the order
parameters (e.g., the $d$-wave above is by itself $d\pm id$), the
ground state can indeed be a nematic state in our case.

To illustrate the appearance of nematic order in the coexistence state,
we write down the a Ginzburg-Landau theory that describes one the four cases
discussed above. The analysis is similar to that done in Ref. \cite{Chichinadze19}
for the case of coexisting SC orders that transform as the $A_{2}$
and $E$ irreps of the $D_{3}$ point group considered in that work.
Here, let $\Delta_{0}$ be the SC order parameter that transforms
as the one-dimensional irrep $A_{1}$ and $\boldsymbol{\Delta}_{E}\equiv\left(\Delta{}_{1},\Delta{}_{2}\right)$,
the SC order parameter that transforms as the two-dimensional irrep
$E_{2}$. Before writing down the SC free-energy expansion, we define
the two subsidiary nematic order parameters $\boldsymbol{\varphi}_{0}$
and $\boldsymbol{\varphi}_{E}$ that arise from bilinear combinations
of the SC order parameters that transform as the $E_{2}$ irrep: 
\begin{align}
\boldsymbol{\varphi}_{0} & =\begin{pmatrix}\Delta_{0}^{*}\Delta_{1}+\Delta_{1}^{*}\Delta_{0}\\
\Delta_{0}^{*}\Delta_{2}+\Delta_{2}^{*}\Delta_{0}
\end{pmatrix}\\
\boldsymbol{\varphi}_{E} & =\begin{pmatrix}\left|\Delta_{1}\right|^{2}-\left|\Delta_{2}\right|^{2}\\
-\Delta_{1}^{*}\Delta_{2}-\Delta_{1}\Delta_{2}^{*}
\end{pmatrix}\label{phi_nematics}
\end{align}

We consider the free energy up to the quartic terms of the SC order parameters.
The free energy for $\Delta_{0}$ is given by the standard form 
\begin{equation}
F_{0}=\frac{\alpha_{0}}{2}|\Delta_{0}|^{2}+\frac{\beta_{0}}{4}|\Delta_{0}|^{4}\ .
\end{equation}

As for the free energy for $\boldsymbol{\Delta}_{E}$, invariance
under $D_{6}$ and U(1) symmetries give: 
\begin{eqnarray}
F_{E} & = & \frac{\alpha_{E}}{2}\left(\left|\Delta_{1}\right|^{2}+\left|\Delta_{2}\right|^{2}\right)+\frac{\beta_{E}}{4}\left(\left|\Delta_{1}\right|^{2}+\left|\Delta_{2}\right|^{2}\right)^{2}\nonumber \\
 &  & -\frac{\gamma_{EE}}{4}\left|\Delta_{1}^{*}\Delta_{2}-\Delta_{1}\Delta_{2}^{*}\right|^{2}\,.
\end{eqnarray}

As we discussed above, if only $\boldsymbol{\Delta}_{E}$ SC order
is present, the SC order parameter is expected to be of the chiral
form $\Delta_{1}\pm i\Delta_{2}$, which breaks time-reversal symmetry
while maintaining the $C_{3z}$ rotational symmetry. This implies
a positive coefficient $\gamma_{EE}>0$.

The coupling between the SC order parameters, $\Delta_{0}$ and $\boldsymbol{\Delta}_{E}$,
up to the quartic order, consists of the following terms 
\begin{eqnarray}
F_{0E} & = & \frac{\lambda}{2}\,|\Delta_{0}|^{2}\left(\left|\Delta_{1}\right|^{2}+\left|\Delta_{2}\right|^{2}\right)\nonumber \\
 &  & +\frac{\gamma_{0E}}{2}\left[\left(\Delta_{0}^{*}\Delta_{1}+\Delta_{0}\Delta_{1}^{*}\right)\left(\left|\Delta_{1}\right|^{2}-\left|\Delta_{2}\right|^{2}\right)\right.\nonumber \\
 &  & \left.-\left(\Delta_{0}^{*}\Delta_{2}+\Delta_{2}^{*}\Delta_{0}\right)\left(\Delta_{1}^{*}\Delta_{2}+\Delta_{1}\Delta_{2}^{*}\right)\right]\nonumber \\
 &  & +\frac{\gamma_{00}}{4}\left[\left(\Delta_{0}^{*}\right)^{2}\left(\Delta_{1}^{2}+\Delta_{2}^{2}\right)+\mathrm{c.c.}\right]\label{Eqn:FreeEneCoupling}
\end{eqnarray}

The total free energy $F\equiv F_{0}+F_{E}+F_{E0}$ can be cast in
a more transparent form by noting that:

\begin{align}
\left|\Delta_{1}^{*}\Delta_{2}-\Delta_{1}\Delta_{2}^{*}\right|^{2} & =-\left(\left|\Delta_{1}\right|^{2}-\left|\Delta_{2}\right|^{2}\right)^{2} \\
 & -\left(\Delta_{1}^{*}\Delta_{2}+\Delta_{1}\Delta_{2}^{*}\right)^{2} +\left(\left|\Delta_{1}\right|^{2}+\left|\Delta_{2}\right|^{2}\right)^{2}\nonumber
\end{align}
and that:

\begin{align}
\left[\left(\Delta_{0}^{*}\right)^{2}\left(\Delta_{1}^{2}+\Delta_{2}^{2}\right)+\mathrm{c.c.}\right] & =\left(\Delta_{0}^{*}\Delta_{1}+\Delta_{1}^{*}\Delta_{0}\right)^{2}\nonumber \\
+\left(\Delta_{0}^{*}\Delta_{2}+\Delta_{2}^{*}\Delta_{0}\right)^{2}- & 2|\Delta_{0}|^{2}\left(\left|\Delta_{1}\right|^{2}+\left|\Delta_{2}\right|^{2}\right)
\end{align}
Then, we have:
\begin{align}
 & F=\left[\frac{\alpha_{0}}{2}|\Delta_{0}|^{2}+\frac{\beta_{0}}{4}|\Delta_{0}|^{4}\right]\nonumber \\
 & +\left[\frac{\alpha_{E}}{2}\left(\left|\Delta_{1}\right|^{2}+\left|\Delta_{2}\right|^{2}\right)+\frac{\left(\beta_{E}-\gamma_{EE}\right)}{4}\left(\left|\Delta_{1}\right|^{2}+\left|\Delta_{2}\right|^{2}\right)^{2}\right]\nonumber \\
 & +\frac{\left(\lambda-\gamma_{00}\right)}{2}\,|\Delta_{0}|^{2}\left(\left|\Delta_{1}\right|^{2}+\left|\Delta_{2}\right|^{2}\right)+\delta F\label{F}
\end{align}
with:

\begin{equation}
\delta F=\frac{1}{4}\left(\begin{array}{cc}
\boldsymbol{\varphi}_{0}^{T} & \boldsymbol{\varphi}_{E}^{T}\end{array}\right)\cdot\left(\begin{array}{cc}
\gamma_{00} & \gamma_{0E}\\
\gamma_{0E} & \gamma_{EE}
\end{array}\right)\left(\begin{array}{c}
\boldsymbol{\varphi}_{0}\\
\boldsymbol{\varphi}_{E}
\end{array}\right)\label{delta_F}
\end{equation}

Note that $F-\delta F$ is insensitive on the relative phases between
the three SC order parameters $\Delta_{0}$, $\Delta_{1}$, and $\Delta_{2}$.
All it does is to determine whether there is a coexistence state in
which both $\Delta_{0}$, $\boldsymbol{\Delta}_{E}$ are simultaneously
non-zero. The symmetry of the coexistence state is determined solely
by the minimization of $\delta F$. In this regard, note that $\gamma_{EE}>0$
favors a relative phase of $\pi/2$ between $\Delta_{1}$ and $\Delta_{2}$,
which corresponds to $\boldsymbol{\varphi}_{E}=0$; $\gamma_{00}>0$
favors a relative phase of $\pi/2$ between $\Delta_{0}$ and $\Delta_{1},\,\Delta_{2}$,
which corresponds to $\boldsymbol{\varphi}_{0}=0$. However, regardless
of the sign of $\gamma_{0E}$, it always favors a state in which the
two nematic order parameters $\boldsymbol{\varphi}_{E}$ and $\boldsymbol{\varphi}_{0}$
are both non-zero and either parallel ($\gamma_{0E}<0$) or anti-parallel
($\gamma_{0E}>0$).

To find out the nature of the coexistence phase, one needs to diagonalize
the matrix in Eq. (\ref{delta_F}). The eigenvalues are given by:
\begin{equation}
\gamma_{\pm}=\left(\frac{\gamma_{00}+\gamma_{EE}}{2\gamma_{0E}}\right)\pm\sqrt{\left(\frac{\gamma_{00}-\gamma_{EE}}{2\gamma_{0E}}\right)^{2}+1}\label{egvalues}
\end{equation}
and the eigenvectors are given by 
\begin{align}
\boldsymbol{\varphi}_{E}=\left[\left(\frac{\gamma_{00}-\gamma_{EE}}{2\gamma_{0E}}\right)\pm\sqrt{\left(\frac{\gamma_{00}-\gamma_{EE}}{2\gamma_{0E}}\right)^{2}+1}\right]\boldsymbol{\varphi}_{0}\label{egvectors}
\end{align}

Because $\gamma_{EE}>0$, the eigenvalue $\gamma_{+}$ is always positive.
However, $\gamma_{-}$ can be negative if $\gamma_{0E}^{2}>\gamma_{00}\gamma_{EE}$.
When this condition is satisfied, the energy is minimized by
condensing the auxiliary nematic order parameter $\boldsymbol{\varphi}_{-}$,
and the coexistence state becomes a nematic superconductor. Note that
$\boldsymbol{\varphi}_{-}$ consists of the two original nematic ``vectors''
$\boldsymbol{\varphi}_{0}$ and $\boldsymbol{\varphi}_{E}$ aligned
parallel to each other, if $\gamma_{0E}<0$, or anti-parallel to each
other, if $\gamma_{0E}>0$. A similar condition for the coexistence
state to be nematic was also found in Ref. \cite{Chichinadze19} for
the case of an $A_{2}$ and an $E$ superconducting states. We emphasize
that additional cubic term in $\boldsymbol{\varphi}_{0}$ and $\boldsymbol{\varphi}_{E}$,
which are allowed by $D_{6}$ symmetry, lower the symmetry of the
nematic order parameter from a two-dimensional vector (i.e. an ``XY''
order parameter) to a 3-state Potts order parameter \cite{Fernandes1,Fernandes2}.

It is interesting to discuss what happens if $\gamma_{0E}^{2}<\gamma_{00}\gamma_{EE}$.
In that case, the $\delta F$ term in the free energy would be minimized
by simultaneously vanishing the two auxiliary nematic order parameters
$\boldsymbol{\varphi}_{0}$ and $\boldsymbol{\varphi}_{E}$. However,
a quick inspection of Eq. (\ref{phi_nematics}) shows that this is
not possible. On the one hand, for $\boldsymbol{\varphi}_{E}$ to
vanish, one needs to impose $\Delta_{1}=\pm i\Delta_{2}$. On the
other hand, $\boldsymbol{\varphi}_{0}$ will only vanish if the relative
phase between $\Delta_{0}$ and $\Delta_{1}$ \emph{and} the relative
phase between $\Delta_{0}$ and $\Delta_{2}$ are $\pm\pi/2$. Thus,
in order for $\boldsymbol{\varphi}_{E}=\boldsymbol{\varphi}_{0}=0$,
the three complex fields $\Delta_{0}$, $\Delta_{1}$, and $\Delta_{2}$
must all have relative phases of $\pm\pi/2$ with respect to each
other, which is impossible. This situation is analogous to geometrically frustrated antiferromagnetism on a triangular lattice.
 The outcome is that the relative phases between $\Delta_{0}$,
$\Delta_{1}$, and $\Delta_{2}$ will be neither $0,\pi$ nor $\pm\pi/2$.
As a result, nematic order will persist even if $\gamma_{0E}^{2}<\gamma_{00}\gamma_{EE}$,
however it will generally be accompanied by time-reversal symmetry-breaking
\cite{Chichinadze19}. In this sense, this order is more appropriately referred to as a \emph{chiral-nematic} order.

The above analysis focused on the case where the two coexisting SC
order parameters were $A_{1}$ and $E_{2}$. Similar expressions
can be derived for the other cases discussed in Table \ref{tab:summary}.
Let us denote the $E_{1}$ or $E_{2}$ SC order parameters generically
as $\boldsymbol{\Delta}_{E}\equiv\left(\Delta{}_{1},\Delta{}_{2}\right)$,
and the $A_{2}$, $B_{1}$, or $B_{2}$ SC order parameter, $\Delta_{0}$.
The free energy acquires the same form as Eqs. (\ref{F}) and (\ref{delta_F}),
with $\boldsymbol{\varphi}_{E}$ still defined by Eq. (\ref{phi_nematics}),
but $\boldsymbol{\varphi}_{0}$ acquiring different functional forms.
In particular, we have 
\begin{align}
\boldsymbol{\varphi}_{0}^{(A_{1},E_{2})/(B_{1},E_{1})} & =\begin{pmatrix}\Delta_{0}^{*}\Delta_{1}+\Delta_{1}^{*}\Delta_{0}\\
\Delta_{0}^{*}\Delta_{2}+\Delta_{2}^{*}\Delta_{0}
\end{pmatrix}\\
\boldsymbol{\varphi}_{0}^{(A_{2},E_{2})/(B_{2},E_{1})} & =\begin{pmatrix}-\Delta_{0}^{*}\Delta_{2}-\Delta_{2}^{*}\Delta_{0}\\
\Delta_{0}^{*}\Delta_{1}+\Delta_{1}^{*}\Delta_{0}
\end{pmatrix}
\end{align}

Regardless of the definition of $\boldsymbol{\varphi}_{0}$, the
condition for a nematic superconductor to take place in all these
coexistence phases is the same as before, $\gamma_{0E}^{2}>\gamma_{00}\gamma_{EE}$.

Although in this section we studied the emergence of nematic order
when the system has already developed long-range SC order, it is worth
emphasizing that nematic order may be established close to but above
the SC transition temperature, as long as the SC fluctuations are
strong enough. In this case, the nematic order is understood as a
vestigial SC order. A detailed analysis of the vestigial order requires one to go beyond mean-field theory and rewrite the free energy in terms of composite order parameters~\cite{Vestigial}, which we leave for future work.

\section{Summary}

\label{sec:Summary}

%\textcolor{red}{{[}RMF: I did some rewriting in these paragraphs, please check{]}} 
In this work we analyzed an SU(4) spin-valley-fermion
model, in which itinerant fermions in a hexagonal two-band system
are coupled to soft fluctuations associated with a large SU(4) symmetry.
Previous theoretical studies have largely focused on either weak-coupling
approaches or strong-coupling approaches. In the latter, which neglects
the kinetic energy, it has been shown that a SU(4) ``ferromagnetic-like''
order arises. In the former, it has been emphasized the importance
of Fermi surface properties such as van Hove singularities, and various
symemtry-breaking intertwined orders have been predicted.

While a self-consistent analysis bridging these two approaches is
remarkably difficult, our phenomonelogical SU(4) spin-valley-fermion
model provides an interesting first step towards this goal. More specifically,
this model assumes that the ferro-SU(4) fluctuations arise from energy
scales much larger than the bandwidth, and treat them as an input
of the theory. The coupling between these ferro-SU(4) fluctuations
and the low-energy fermions has two main effects: on the one hand,
it partially lifts the huge degeneracy associated with the SU(4) manifold.
On the other hand, it gives rise to a rich landscape of superconducting
phases with non-trivial topology, nodes, and broken time-reversal
symmetry. Superconducting nematicity, as signaled by the spontaneous
breaking of the lattice rotational symmetry inside the superconducting
state, emerges quite naturally due to the near degeneracy of the competing
pairing states, despite the fact that none of the superconducting
states are by themselves nematic.

Remarkably, these results are robust despite the lack of a detailed
knowledge of the Fermi surface of twisted bilayer graphene, as they
only depend on general features of the Fermi surface, namely, the
valley hot-spots and the van Hove hot-spots. In this regard, the SU(4)
spin-valley-fermion model is the TBG counterpart of the widely-studied
SU(2) spin-fermion model usually applied to cuprate superconductors.
The main differences are the existence of valley degrees of freedom
and the condensation of the SU(4) order parameter in a ``ferromagnetic-like''
configuration, i.e. an ordered state with zero wave-vector. Theoretically,
an interesting issue for future investigations is about the quantum
critical properties of the ferro-SU(4) model, and how they may be
manifested in thermodynamic and transport properties. Experimentally,
the extent to which these results apply directly to TBG remains an
open question that certainly deserves further studies. In particular,
while ferromagnetic order has been observed in quarter-filling TBG,
it is presently unclear whether the half-filled insulating state is accompanied
by any broken symmetry.

\begin{acknowledgments}
We thank A. Chubukov, D. Chichinadze, L. Classen, S. Kivelson, and O. Vafek for fruitful
discussions. Y.W. is supported by the startup funds at University
of Florida. J.K. is supported by Priority Academic Program Development
(PAPD) of Jiangsu Higher Education Institutions. R.M.F. was supported
by the U. S. Department of Energy, Office of Science, Basic Energy
Sciences, Materials Sciences and Engineering Division, under Award
No. DE-SC0020045. We thank the hospitality of the Aspen Center for
Physics, supported by NSF PHY-1066293, where this work was initiated.
\end{acknowledgments}

%Unused bibitems

%\bibitem{Kennes18} D. M. Kennes, J. Lischner, and C. Karrasch, Phys.
%Rev. B \textbf{98}, 241407(R) (2018).
%
%\bibitem{Bascones19} J. M. Pizarro, M. J. Calder\'on, and E. Bascones,
%Journal of Physics Communications \textbf{3}, 035024 (2019).
\end{document}